\documentclass{article}
\usepackage{graphicx}
\usepackage{xcolor}
\usepackage{parskip}
\usepackage{enumitem}%
\usepackage{comment}
\usepackage{caption}
\usepackage{float}
\usepackage{booktabs}
\usepackage{color,soul}
\usepackage{tabularx}
\usepackage{amsmath}
\usepackage{bm}
\usepackage{authblk}
\usepackage[numbers,sort&compress]{natbib}

\usepackage[a4paper, margin=1in]{geometry}

\usepackage{blindtext}

\usepackage{titlesec}
\titleformat*{\section}{\bfseries}

\begin{document}

\title{\huge{Myofibroblasts slow down defect recombination dynamics in mixed cell monolayers}}
\vspace{0.1in} 
\author[1,2]{Zhaofei Zheng}
\author[1]{Yuxin Luo} %
\author[1,3]{Juan Chen}
\author[1,*]{Yimin Luo}
\affil[1]{Department of Mechanical Engineering, Yale University, New Haven, CT}
\affil[2]{Present address: Department of Physics, Carnegie Mellon University, Pittsburgh, PA}

\affil[3]{Present address: University of Maryland Medical Center, Baltimore, MD}
\affil[*]{Corresponding author: yimin.luo@yale.edu}
\date{}
\maketitle

\begin{abstract}

Cellular organization and mechanotransduction pathways are crucial regulators of tissue morphogenesis, whereas their dysregulation contributes to pathologies. Overactive myofibroblasts are key drivers of fibrosis, yet how their presence alters collective cellular ordering remains unclear. Owing to steric interactions, elongated cells exhibit local order. Topological defects, where alignment is disrupted, have been postulated to serve as mechanical centers. In this study, we examine how incorporating slower moving myofibroblast phenotype impacts defect relaxation in monolayers consisting of co-cultured fibroblasts and myofibroblasts. 
In this system, myofibroblasts act as the less active component.
Increasing their fraction increases disorder strength and slows defect recombination. On microgrooved surfaces, higher myofibroblast concentrations lead to worse alignment, suggesting single-cell mechanosensing and cell-cell interactions act jointly. 
Furthermore, we found that myofibroblasts preferentially localize at negatively charged -$\frac{1}{2}$ defects, whereas fibroblasts localize at +$\frac{1}{2}$ defects. Consequently, the slowdown of recombination dynamics can be partially attributed to myofibroblasts’ preferential association with the less mobile -$\frac{1}{2}$ defects, increasing local friction and impeding defect mobility. This localization may also reduce compressive stress on myofibroblasts, as indicated by immunofluorescence of a downstream mechanotransducer. This work provides insights into possible connections between topological defects and cell motility in mixed phenotype monolayers.

\end{abstract}



\section{Introduction}

Liquid crystals (LCs) underlie technological advancement, exemplified by LC displays. Many biological systems, from cytoskeletons to lipid bilayers \cite{stewart2003liquid,stewart2004liquid}, also exhibit LC phases.  A growing body of literature draws parallels between cell ordering and nematic LC, where the orientation of an individual cell is represented by a headless vector (the ''director''), and steric interaction between cells causes neighboring cells to co-align \cite{endresen2021topological, balasubramaniam2021investigating, kawaguchi2017topological, yashunsky2022chiral}. In addition, cells also align parallel with physical boundaries \cite{bade2018edges, duclos2014perfect}, akin to LC anchoring, and exhibit deformations analogous to classical splay and bend modes \cite{murali2025splay}. As density increases, elongated cells undergo a disorder-to-order transition \cite{luo2022cell}. Cell–cell alignment amplifies single-cell mechanotransduction, suggesting a collective mechanism for sensing substrate anisotropy, until the system eventually becomes arrested at high density \cite{parmar2026proliferating}.

Cell alignment is typically imperfect in a monolayer; the order can be quantified by the order parameter $S$, which ranges from 0 (random) to 1 (perfect alignment). Alignment breaks down at topological defects, regions within the cell assembly where order becomes frustrated ($S$ = 0). In 2D, defects are characterized by a topological ''charge'' measuring the angle by which the director field rotates as one encircles the defect.  Originally considered an academic novelty, defects were later thought to contribute to tissue maintenance, and their movement can provide insights into the nature of the underlying forces \cite{saw2017topological,kawaguchi2017topological,balasubramaniam2022active}.
Over time, cells accumulate at $+\frac{1}{2}$ defects and become depleted at -$\frac{1}{2}$ defects. Due to compressive stresses concentrating at the $+\frac{1}{2}$ defects, cells can form mounds \cite{kawaguchi2017topological}, and undergo elevated rates of apoptosis and extrusion \cite{saw2017topological}. This analogy could be extended to 3D, paving the way for controlled force generation and shape transformation \cite{huang2024cell,duclos2017topological}. Perhaps most strikingly, topology-driven morphogenesis can operate at the scale of entire organisms; a well-known example is the role of topological defects in establishing a robust body plan in Hydra \cite{maroudas2021topological}.

Nematic LC order has been uncovered in an ever-growing list of cell types \cite{doostmohammadi2021physics}, from muscle \cite {martella2019liquid}, breast cancer \cite{saw2017topological}, to retinal cells \cite{duclos2018spontaneous}.
While the LC ordering in cells is now well-recognized, leveraging the active nematics framework to gain insights into development and disease progression remains a recent and rapidly emerging area of research.  Topological defects have been shown to guide the self-assembly of stromal cells into cartilage tissues \cite{makhija2024topological}. In the mesothelial layer, defects direct tissue flow, with regions of net inward flow inhibiting tumor cell clearance \cite{zhang2021liquid}. Aligned layers of cancer-associated fibroblasts act as defensive barriers, whereas topological defects represent weak spots that can be exploited for cancer cell dissemination \cite{jacques2023aging}. 
More recently, brain cancer cells (gliomas) have been found to exhibit large-scale ordering, with the degree of this organization correlating to the tumor aggressiveness \cite{argento2025three,lowenstein2025oncostreams}. 

One underexplored context is how the LC order is mediated in a system that consists of two distinct cell phenotypes. Phenotype refers to organisms with the same genetic composition, but exhibiting different physical characteristics owing to their interaction with environmental factors. 
For example, when fibroblasts, the connective tissue cells,  encounter stiff environments \cite{cabrera2012mechanical,wipff2007myofibroblast} or chemical signals \cite{pakyari2013critical}, they turn into a strongly contractile but less mobile phenotype \cite{d2025substrate}, myofibroblasts, to facilitate wound healing. These same myofibroblasts are also known drivers for fibrosis \cite{juhl2020dermal}. 
During normal tissue healing, myofibroblasts undergo apoptosis, or programmed cell death, upon wound closure \cite{tomasek2002myofibroblasts}. However, during fibrotic disease progression, the activated myofibroblasts surprisingly evade apoptosis \cite{hinz2020evasion}.  While single-cell force profiling and fluorescence staining can provide valuable insights \cite{yang2021quantitative}, they may not fully capture the complexity of these multicellular dynamics due to their limited sample size and static nature. Physiological processes such as wound healing, maintenance, and morphogenesis depend on coordinated forces from heterogeneous populations, while loss of tissue organization often indicates underlying inflammatory conditions \cite{ishida2020basket}. 
There is an unmet need for simplified, predictive in vitro models that can clarify the mechanisms linking cell phenotypes to tissue organization and enable quantitative monitoring of fibrosis without sampling-related biases. 

In this work, we study how the fraction of myofibroblasts in the population influences the dynamics of monolayer organization. 
Here, frictional interactions dominate over hydrodynamic momentum transport through the surrounding fluid. The collective dynamics of the cells are governed by local cell–cell interactions, alignment of cell shapes, and active stresses generated by the cells, rather than being transmitted by fluid flows. Such systems are often called a ``dry'' active nematic \cite{chate2020dry}. In this setting, we must distinguish between motility and contractility. Because the cells are not coupled to flow or interacting through collagen networks, myofibroblasts represent the less active phenotype, given that they move more slowly.

We create an in vitro system consisting of varying ratios of fibroblasts and myofibroblasts to mimic the progressive nature of fibrosis. 
Using time-lapse live-cell microscopy and immunofluorescence, we study topological defects that arise naturally in dense monolayers and when cells are cultured on topographically patterned substrates. Myofibroblasts adopt an elongated morphology in response to surrounding fibroblasts. Together, these two phenotypes form an active nematic layer, 
where elasticity dominates over activity.
Our study reveals that defect relaxation slows down with increasing myofibroblast concentration, reminiscent of systems with quenched disorder, with the myofibroblast fraction ($\Phi_\mathrm{MF}$) acting as a tunable disorder strength. Using a substrate with unidirectional orientation, we demonstrate that monolayers containing more myofibroblasts exhibit decreased alignment order. 
Finally, we find that distinct phenotypes develop affinity for defects of different topological charges. Less active myofibroblasts exploit defects in long-range LC organization by avoiding the more motile, positively-charged defects. Together, these findings suggest that defects create a distinct mechanical environment and illuminate the feedback loop between defect stabilization and local myofibroblast enrichment.local myofibroblast enrichment.

\section{Results}
\subsection{Tracking defect dynamics in progressive disease model}

To characterize typical dermal compositions, dermal biopsy samples were obtained from healthy donors and individuals with systemic sclerosis (SSc, more details can be found in SI, Section S1), where fibrosis is a key symptom. Immunofluorescence staining of tissue samples from patients with SSc shows more activated myofibroblasts compared to a healthy control (Fig. S1a). The biomarker $\alpha$-SMA is overexpressed in patients with fibrotic disease, while remaining largely absent in healthy donors. We estimate that cells expressing $\alpha$-SMA could be as high as 50-70\% in SSc patients  (Fig. S1b). Their presence can impair dermal function, causing joint stiffness and increased scarring. In dense layers, however, the two phenotypes are difficult to distinguish based on morphology alone, and their mixed presence and proximity suggest their ongoing crosstalk during fibrotic disease progression.

\begin{figure}[htb!]
\begin{center}
    \includegraphics[width=0.75\textwidth]{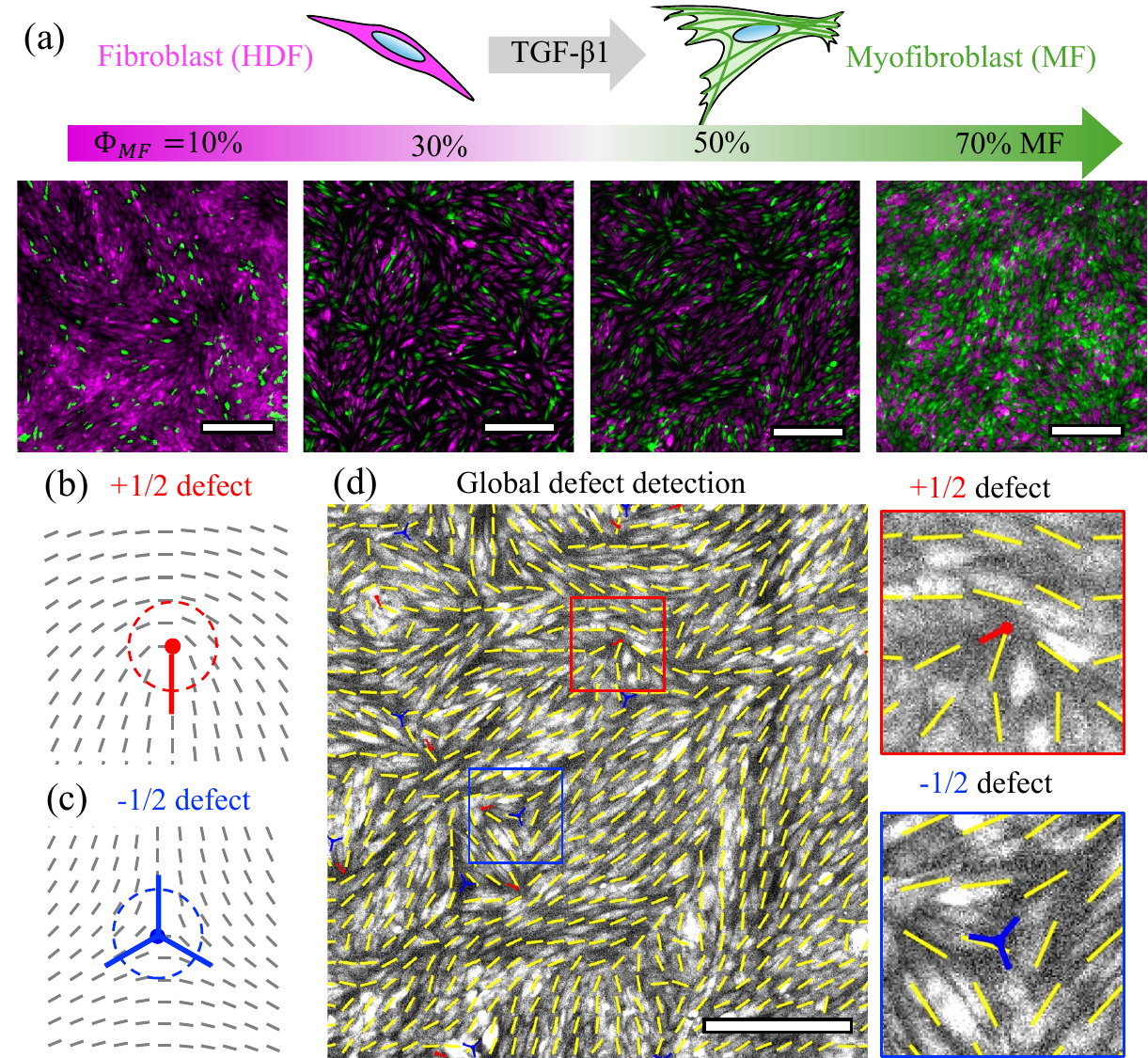}
\end{center}
  \caption{(a) Fluorescent micrographs of mixed fibroblast (magenta) and myofibroblast (green) monolayers, where the percentage of myofibroblasts is systematically increased, mimicking the progression of fibrotic diseases. (b-c) Schematics of topological defects with charges $\pm\frac{1}{2}$. (d) Defect detection in a real tissue monolayer, where the detected director field is plotted as an overlay in yellow, the cores of the defects are denoted in red ($+\frac{1}{2}$ defects) and blue ($-\frac{1}{2}$ defects), respectively. The inset shows a zoomed-in view of both types of defects. The scale bars are 500 $\mu$m. }
\label{fig: defect_detection}
\end{figure}

To better isolate and control myofibroblast concentration in the sample, we construct a simplified in vitro model with mixed phenotypes to mimic progressive disease states. We induce myofibroblasts (MFs) in vitro (Fig. \ref{fig: defect_detection}a and Fig. S2) from adult human dermal fibroblasts (HDFs), by adding transforming growth factor-$\beta$1 (TGF-$\beta$1, 10 ng/mL), a key mediator released by platelets during wound healing, to mimic the chemical environment at the wound site. Both induced and untreated cells retain their phenotypes in media with a reduced concentration of TGF-$\beta$1 (2 ng/mL, Fig. S2). After labeling, the induced cells are counted and mixed with untreated fibroblasts. This procedure results in a well-defined fibroblast (magenta) to myofibroblast (green) ratio, and each phenotype is distinguished by different channels (Fig. \ref{fig: defect_detection}a). Throughout this study, we avoid using DAPI for live cell imaging to minimize its detrimental effects on the cells. DAPI staining is only used after cell fixation for easier segmentation. We quantify the fraction of myofibroblasts, $\Phi_\mathrm{MF}$, using three independent approaches in experiments: cell segmentation with Cellpose, nucleus detection using TrackMate, or estimation from the areal fraction after binarizing individual channels. All three methods yield comparable  $\Phi_\mathrm{MF}$ (see Section S3 and Fig. S3 for details).

Consistent with literature \cite{d2025substrate}, we find the myofibroblasts move more slowly than fibroblasts (Fig. S4), confirming that they are the less active phenotype. At high density ($\rho_\mathrm{cell} \approx$ 750 cells mm$^{-2}$), the monolayer behaves as a relaxing solid, exhibiting long-range, correlated motion (Videos S1). In contrast, at lower concentrations ($\rho_\mathrm{cell} \approx$ 500 cells mm$^{-2}$), individual cells move back and forth without coherent dynamics (Videos S2). Particle image velocimetry (PIV) and velocity correlation (Section S5 and Fig. S5) reveal that monolayers at  $\rho_\mathrm{cell} \approx$ 500 cells mm$^{-2}$ have a velocity correlation length comparable to a single cell ($\approx 100\ \mu\text{m}$). On the other hand, at $\rho_\mathrm{cell} \approx$ 750 cells mm$^{-2}$, the velocity correlation length increases substantially to hundreds of $\mu\text{m}$.
We are primarily interested in how embedding myofibroblasts affects global relaxation dynamics. Thus, we focus on $\rho_\mathrm{cell} \approx$ 750 cells mm$^{-2}$ in the subsequent sections. At this density, friction dominates over activity, and proliferation is largely suppressed. 

Fibroblast-myofibroblast monolayers exhibit nematic order. Steric interactions create aligned domains separated by disordered regions where topological defects emerge. 
In 2D, stable defects appear as either +$\frac{1}{2}$ defect, also known as a ''comet'' (Fig. \ref{fig: defect_detection}b), or $-\frac{1}{2}$ defect, also known as a ''trefoil'' (Fig. \ref{fig: defect_detection}c). 
Our goal is to identify the orientation field and track defect dynamics from microscopy videos. To this end, we first apply the structure tensor method \cite{kawaguchi2017topological}, with the results shown as the yellow overlay (Fig. \ref{fig: defect_detection}d). Then we identify defects using a topology-preserving algorithm \cite{tan2019topological}, which automatically identifies the $\pm\frac{1}{2}$ defects, as illustrated in the insets of Fig. \ref{fig: defect_detection}d. More details can be found in SI Section S6. Our analysis integrates prior works to suit our specific datasets. We modify the structure tensor method from Ref. \cite{kawaguchi2017topological}, originally developed for phase-contrast microscopy, to work with fluorescence microscopy, because myofibroblasts are relatively flat and therefore exhibit poor contrast in phase contrast mode. We also extend the defect-detecting algorithm from Ref. \cite{tan2019topological}, originally developed for microtubule networks, to cellular data by accounting for the greater discretization errors, and the resulting lower effective resolution, that arise from the larger size of the cells. Finally, we link the defect positions across frames using a particle-tracking scheme, originally developed in Ref. \cite{crocker1996methods}, which yields the full defect trajectories, from which the defect velocities can be computed.

\subsection{Defect relaxation dynamics}

\begin{figure}[htb!]
\begin{center}
    \includegraphics[width=\textwidth]{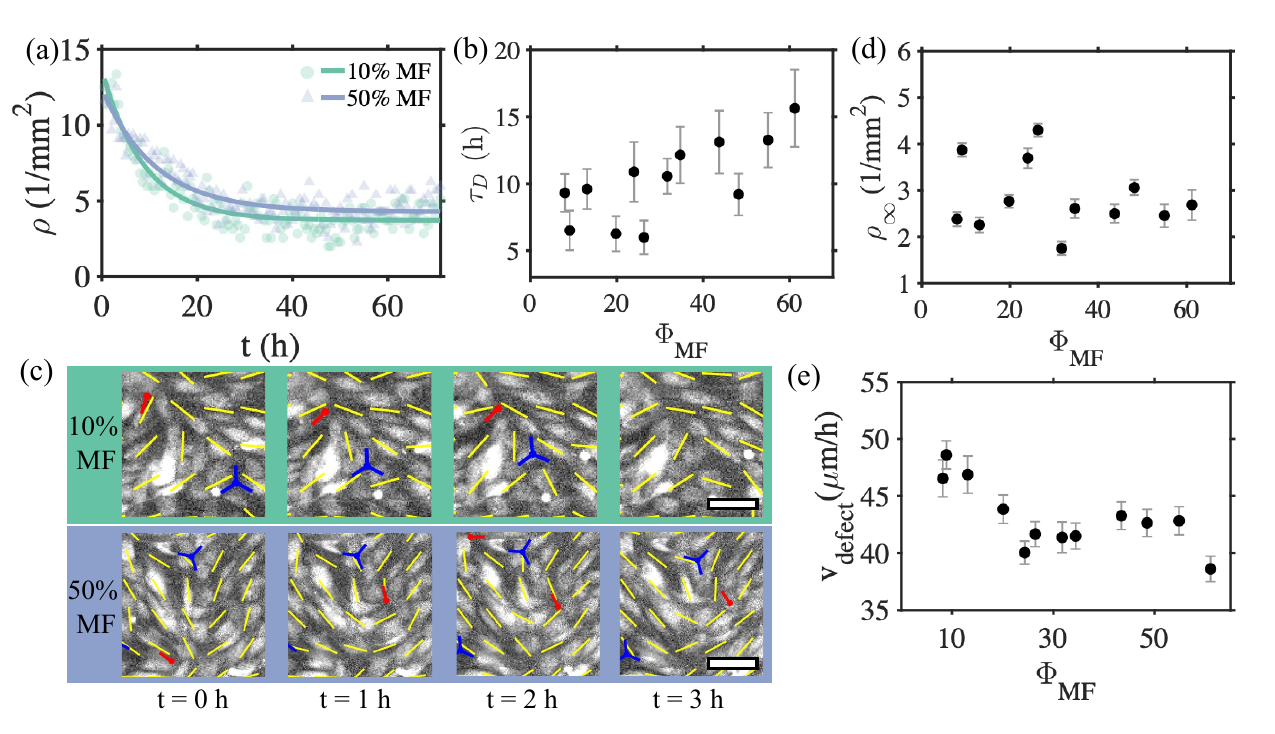}
\end{center}
  \caption{(a) The density of topological defects decreases over time and eventually plateaus at a finite value. Solid lines indicate the exponential fit. (b) Dependence of the decay constant ($\tau_D$) on myofibroblast concentration $\Phi_\mathrm{MF}$.  (c) Time sequence images showing the annihilation of defect pairs with opposite topological charges in both low and high concentrations of myofibroblasts. The scale bars are 100 $\mu$m. (d) Dependence of steady-state defect density, $\rho_\infty$,  as defined in Eq. (3.1), on myofibroblast concentrations $\Phi_\mathrm{MF}$. (e) The average velocity of all defects ($v_\mathrm{defect}$) across myofibroblast concentrations $\Phi_\mathrm{MF}$. }
\label{fig: defect_annilation}
\end{figure}

We observe that the total number of both types of defects $\rho$ decreases over time (Fig. \ref{fig: defect_annilation}a), and eventually reaches a plateau, $\rho \approx$  2 mm$^{-2}$, comparable to earlier observations \cite{sarkar2023crisscross}. For different myofibroblast concentrations, we plot the defect density $\rho$ over time $t$. \textcolor{blue}{The zero time point $t$ = 0 was defined as 6 hours after cell seeding, to allow time for cell attachment.} Additional plots can be found in Fig. S7, and fit an exponential decay:
\begin{equation}
    \rho(t) = (\rho_0-\rho_\infty)e^{-t/\tau_D}+\rho_\infty,
\end{equation}
where $\tau_D$ and $\rho_\infty$ are the decay constant and steady-state defect numbers, respectively. Monolayers with different compositions of the two phenotypes exhibit distinct dynamics. We find that $\tau_D$ increases for monolayers with increasing $\Phi_\mathrm{MF}$ (Fig. \ref{fig: defect_annilation}b). Correspondingly, we also find that defect recombination occurs notably more slowly in myofibroblast–fibroblast mixtures with increasing $\Phi_\mathrm{MF}$ (Fig. \ref{fig: defect_annilation}c). In contrast, the plateau defect density, $\rho_\infty$ (Eq. 3.1), remains unchanged (Fig. \ref{fig: defect_annilation}d).  $\rho_\infty$ is comparable to literature values \cite{sarkar2023crisscross}, and is consistent with a system with low activity \cite{zhang2021spatiotemporal}. 
While occasional spurious defect pair creation events are detected, annihilating processes dominate over creation in our system. This is because fibroblasts and myofibroblasts are both contractile. As a result, while defect pairs are continuously created and annihilated, annihilation events are typically favored compared to extensile systems. We observe an approximately 25\% reduction in the average defect velocity, as myofibroblast concentration in the monolayer composition increases by about six-fold (Fig. \ref{fig: defect_annilation}e).

We find that the slowdown of the defect recombination dynamics persists even when myofibroblasts constitute the majority of the sample ($\Phi_\mathrm{MF}>$ 50\%). Although isolated myofibroblasts typically exhibit a more rounded morphology (Fig. S2), 
they become polarized in the presence of fibroblasts. Under these conditions, the two phenotypes co-migrate, collectively exhibiting local alignment and robust nematic order.   
Myofibroblasts contribute to the overall dynamics not only as potential sources of disorder but also as active constituents that shape the collective motion. Consequently, the monolayer should be analyzed as an integrated system, rather than as two distinct subpopulations, to fully capture the continuous steric and mechanical interactions between the two phenotypes.

Owing to myofibroblasts' slower dynamics, they behave as quasi-quenched heterogeneities at short timescales, hindering defect motion and recombination. This system recapitulates some features of quenched nematics \cite{kumar2020active,kumar2022active}. In such systems, random field disorder partitions the material into uncorrelated domains and hinders the kinetics of structural reorganization \cite{kumar2020active}. 
As the disorder strength increases, both defect recombination dynamics and long-range order are suppressed. In our system, myofibroblasts act as the source of disorder, and increasing myofibroblast concentration increases the disorder strength.

Both cell velocities and defect velocities decrease with increasing myofibroblast concentration (Fig. \ref{fig: defect_annilation}c, Fig. S5). Notably, topological defects often move much faster than the surrounding material, as their motion is governed by the system’s energy landscape and structural configuration rather than by bulk material transport. Moreover, despite the reduction in average speed, velocity correlations change only minimally with increasing myofibroblast $\Phi_\mathrm{MF}$.
This observation is consistent with a recent computational study \cite{kinoshita2025active} that reported that velocity correlations are less sensitive to the strength of noise. Furthermore, the study predicts that the random field arrests the director into a static, disordered configuration. Because velocity is coupled to the director, these frozen distortions continuously generate persistent cell motion along the local director, with cells traveling in an antiparallel fashion, similar to what is observed in Ref.\cite{leclech2022topography}.

\subsection{Alignment with an uniaxial imposed orientation}

\begin{figure}[htb!]
    \centering
    \includegraphics[width=0.95\linewidth]{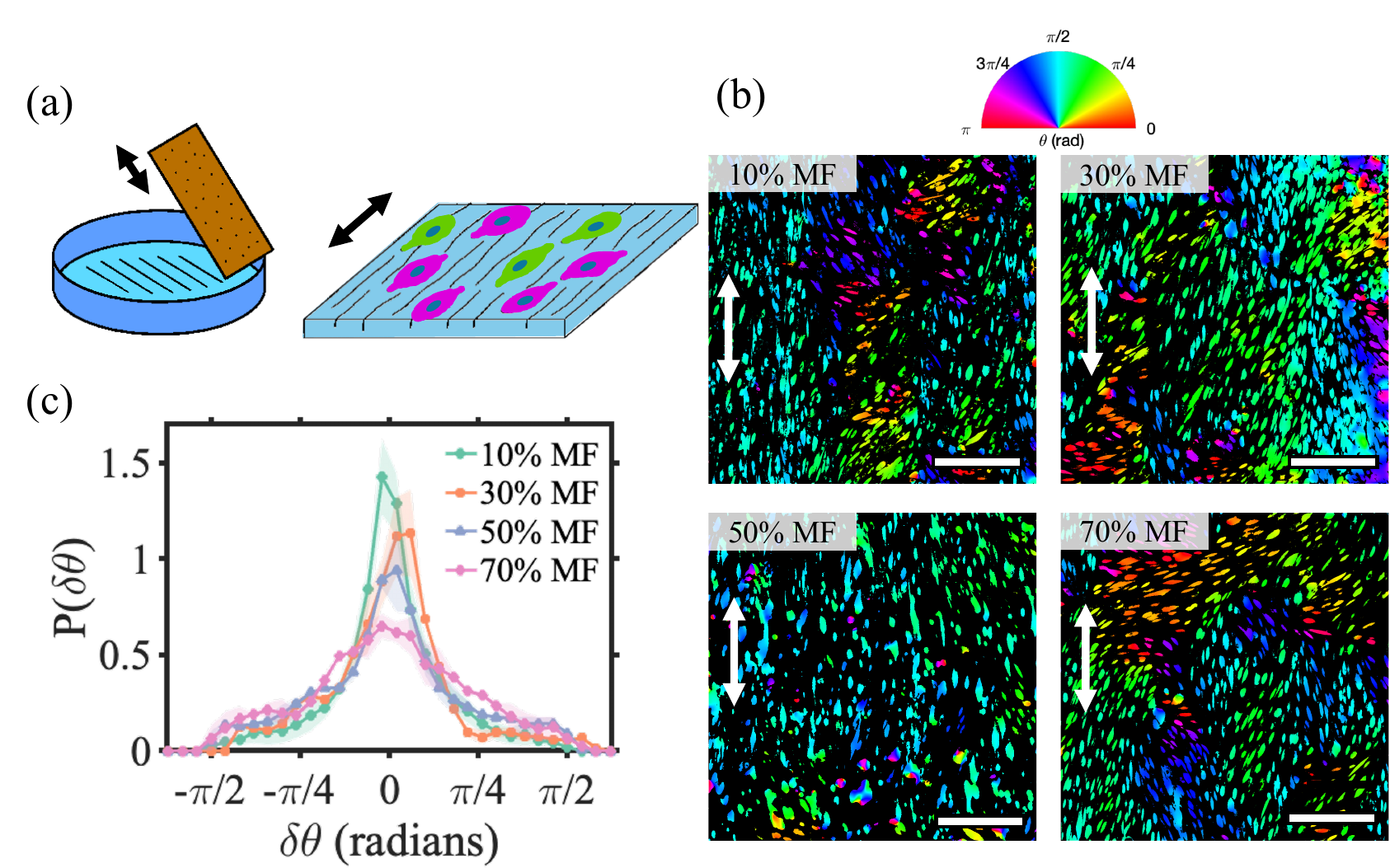}
    \caption{(a) Schematic of the rubbing protocol used to align cells, with cells orienting along the rubbing direction (denoted by double-sided arrows). (b) Representative microscopy images that color-code the orientation of the cells. The scale bars are 500 $\mu$m. (c) Polar distribution of director field angles. Variation angle $\delta \theta$. Shaded region denotes uncertainty average from at least N = 3 different repeats.
    }
    \label{fig:substrates}
\end{figure}

Surface topography can impose strong orientational guidance on fibroblasts and myofibroblasts through contact guidance \cite{babakhanova2020cell,chen2022nematic,pot2010nanoscale}. One study directly compared the sensitivity of fibroblasts versus myofibroblasts to the dimension of topographic features, but the comparison was carried out for individual cells \cite{pot2010nanoscale}. Does increasing $\Phi_\mathrm{MF}$ in the monolayer also impede its ability to align with the imposed orientation? To answer this question, the two phenotypes are premixed at various $\Phi_\mathrm{MF}$ and seeded onto the substrates that are mechanically rubbed with sandpaper (Fig. \ref{fig:substrates}a). This method establishes directional cues that are spaced at length scales larger than the cells (Fig. S8). This spacing ensures the guidance in cell orientation is not imposed on individual cells, allowing cell–cell and cell–substrate interactions to collectively establish global alignment. Cells were seeded at $\rho_\mathrm{cell} \approx 750~\mathrm{cells~mm^{-2}}$, and then fixed after 96 hours, to allow sufficient time for the cells to reach the final arrested state. Cell alignment is characterized by the 2D order parameter $S = \langle2\cos^2\theta-1\rangle$, where $\theta$ denotes the deviation of the cell angle (Fig. \ref{fig:substrates}b) extracted from the preferred orientation (denoted by the double-sided arrow in Fig. \ref{fig:substrates}a). 
The order parameter $S$ decreases from around 0.6 to around 0.3 as the $\Phi_\mathrm{MF}$ increases from 10\% to 70\%. The slowdown in defect recombination dynamics similarly leads to a decrease in $S$ when cells are seeded on a unidirectionally oriented substrate.
 Additionally, we plot the fluctuation of angular distributions $P(\delta\theta)$ in Fig. \ref{fig:substrates}c. Consistent with what is predicted in Ref. \cite{kumar2020active}, the $P(\delta\theta)$ broadens with increasing $\Phi_\mathrm{MF}$, reaffirming myofibroblasts' role as the source of quenched disorder in our system, which acts to inhibit reorganization. If alignment were enforced at the single-cell level, the angle distribution for all samples would appear identical, i.e. narrowly peaked at zero, as each cell would be perfectly aligned. Instead, only a subset of cells align via contact guidance from the features, while the remainder organize according to their interactions with their neighbors. Cells are allowed to organize for four days until the structures are arrested, after which samples with higher myofibroblast fractions exhibit reduced ordering due to slower reorganization. Therefore, the rubbed substrate may also serve as a quantitative and accessible platform for monitoring the fraction of myofibroblasts in the sample, which is an indicator of fibrotic disease progression. 

\subsection{Different phenotypes develop different affinities to different defects}

\begin{figure*}[t!]
    \centering
    \includegraphics[width=0.95\linewidth]{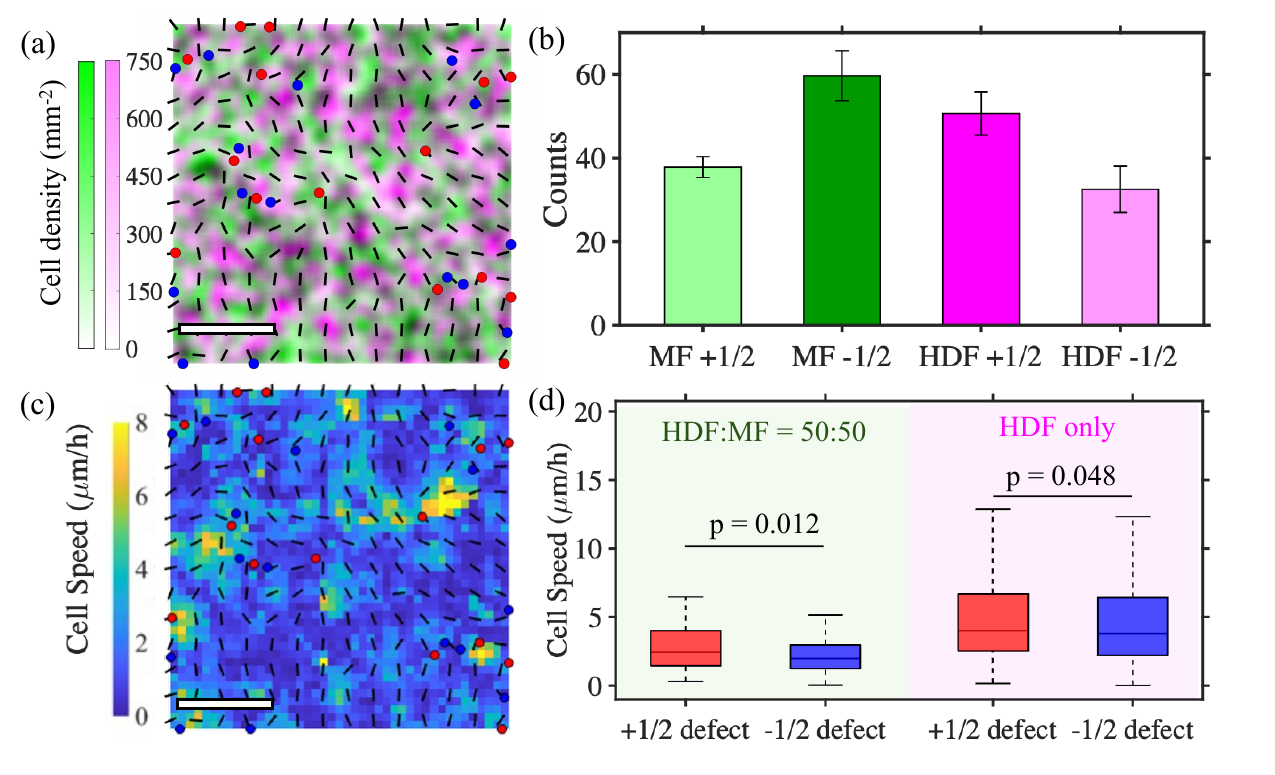}
    \caption{(a) Combined density plot for fibroblasts (magenta) and myofibroblasts (green). (b) Counts of fibroblasts and myofibroblast-enriched regions that coincide with $\pm \frac{1}{2}$ defects (n = 211 defects). (c) Spatial velocity map obtained from particle image velocimetry.  (d) Box and whisker plots of velocities at $\pm \frac{1}{2}$ defect sites for fibroblast-only (left, from n$_{+1/2}$ = 388 defects and n$_{-1/2}$ = 390 defects) and 50:50 myofibroblast-fibroblast mixture (right, from n$_{+1/2}$ = 109 defects and n$_{-1/2}$ = 121 defects). The scale bars are 500 $\mu$m in panels (a) and (c).}
    \label{fig:defect_preference}
\end{figure*}

To understand why higher myofibroblast levels impede organization in monolayers, we compare defect locations with myofibroblast-rich regions. For consistency, we focus on a monolayer with $\Phi_\mathrm{MF} \approx$ 50\% myofibroblast. Fluorescent channels were segmented using CellPose \cite{stringer2021cellpose}. The densities of both phenotypes are merged and overlapped with the director field and the defect (Fig. \ref{fig:defect_preference}a). This reveals that myofibroblast-rich regions (green) coincide more with $-\frac{1}{2}$ defects (blue), while fibroblast-rich regions (magenta) coincide more with $+\frac{1}{2}$ defects (red). To assess whether this trend is general, we present cumulative statistics showing the preference for positively or negatively charged defects for each phenotype in Fig. \ref{fig:defect_preference}b, tallied from n = 211 defects using a custom algorithm paired with manual verification. The algorithm counts cells and calculates the difference between MF and HDF in the vicinity of each defect. This difference is then normalized by the total cell count. If the normalized difference exceeds a predefined threshold, one phenotype is classified as more enriched than the other. Importantly, the number of cells assigned to each category remains stable across a wide range of threshold values. Details of the code and the benchmarking procedures are provided in SI Section S9. The average and standard deviation of the counts for different threshold values are shown in Fig. \ref{fig:defect_preference}b. Our analysis indicates that myofibroblasts preferentially collocate with negatively charged -$\frac{1}{2}$ defects (61\% of the time), whereas fibroblasts tend to associate with positively charged +$\frac{1}{2}$ defects (61\% of the time). 

We hypothesize that the defect dynamics presented in Fig. \ref{fig: defect_annilation} are highly correlated with their compositions. Defects with charge $+\frac{1}{2}$ can self-propel whereas defects with charge $-\frac{1}{2}$ are often more stagnant \cite{duclos2020topological,balasubramaniam2021investigating,sarkar2023crisscross,giomi2014defect}. The velocity field extracted by PIV shows $+\frac{1}{2}$ defects tend to move faster (Fig. \ref{fig:defect_preference}c, yellow regions). Further analysis reveals that regions in the vicinity of -$\frac{1}{2}$ defects travel at statistically significantly lower velocities compared with those surrounding +$\frac{1}{2}$ defects (Fig. \ref{fig:defect_preference}d). Thus, it appears that myofibroblasts slow down global reorganization dynamics through their association with the slower-moving $-\frac{1}{2}$ defects. Previous studies showed that +$\frac{1}{2}$ defects tend to migrate \cite{zhang2021spatiotemporal} and accumulate \cite{chaithanya2024transport} within regions or phases of higher activity. In our system, the fibroblast microphase corresponds to the high-activity region, characterized by increased motility, and they coincide with +$\frac{1}{2}$ defects. In contrast, the myofibroblast microphase exhibits slower dynamics, and tends to coincide with -$\frac{1}{2}$ defects. 

We repeat the same experiment for fibroblast-only monolayers at the same cell density and tracked velocities at $\pm\frac{1}{2}$ defects. We find that the velocities near $\pm\frac{1}{2}$ defects also differ even when cells are homogeneous, due to defects' distinct symmetries, as shown in Fig. 4d. However, the velocity ratio in a pure-fibroblast monolayer is smaller than that observed in the fibroblast–myofibroblast mixture. This is confirmed by performing a one-tailed bootstrap test to verify that $\log\left(\frac{v_\mathrm{+1/2
}}{v_\mathrm{-1/2}}\right)_\mathrm{HDF\!-\!MF}$ is greater than $\log\left(\frac{v_\mathrm{+1/2
}}{v_\mathrm{-1/2}}\right)_\mathrm{HDF}$ ($p<$ 0.0001). In the HDF-MF system, the relative enrichment of myofibroblasts appears to introduce additional friction that slows the motion of $-\frac{1}{2}$ defects.
These observations are consistent with a system exhibiting increased friction \cite{doostmohammadi2018active} and reduced motility, hallmarks of the myofibroblast phenotype. 
 
\subsection{Defect pinning on LCE fiber substrates}

Mobile defects discussed so far move at substantial speeds ($\sim$10s of $\mu$m/h), making cell-defect interactions challenging to follow. Therefore, lastly, we turn to pinned defects, which cannot relax or move. As discussed so far, we find that the mixture of fibroblasts and myofibroblasts can align with the imposed orientation, and their spatial arrangement is influenced by the mechanical forces and the distinct affinities of each cell type for different defects. 

\begin{figure}[htb!]
    \centering
\includegraphics[width=0.9\linewidth]{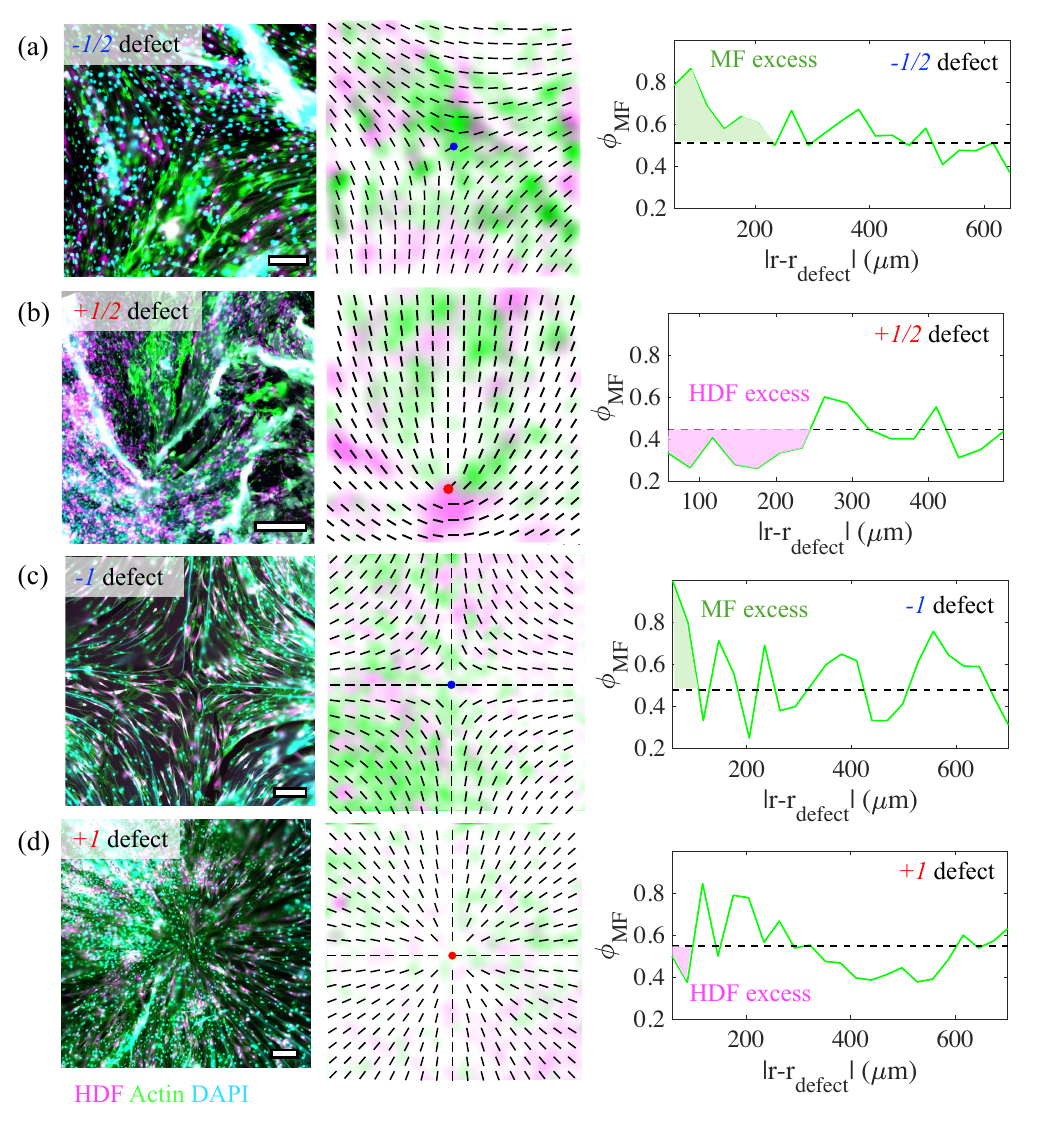}
    \caption{Cells cultured on LCE fibers with patterns of topological charges (a) $-\frac{1}{2}$, (b) $+\frac{1}{2}$, (c) $-1$, and (d) $+1$. Column 1 shows the merged fluorescent micrographs with HDF (magenta), actin (green), and nuclei (blue). The scale bars are 200 $\mu$m. Column 2 shows the cell density map overlaid with the underlying director field and the defects. Column 3 shows the local myofibroblast concentration $\phi_\mathrm{MF}$ within an annulus at a distance $|\mathbf{r}-\mathbf{r}_\mathrm{defect}|$ from the center of the defect. The dashed lines denote the sample average of myofibroblast concentration. The shaded area denotes the range where one phenotype is in excess near the defect core.}
    \label{fig:patterned_defects}
\end{figure}

We achieve defect pinning by creating liquid crystal elastomer (LCE) fiber patterns of topological charge $\pm \frac{1}{2}$ following procedures in Refs. \cite{chen2022nematic,huang2024cell} (Fig. \ref{fig:patterned_defects}a,b, more details can be found in Materials and Methods), coating them with collagen, and seeding them with a $\Phi_\mathrm{MF} \approx$ 50\% mixtures. Collagen coating promotes cell adhesion and prevents delamination, especially at high cell density, but it does not change the surface morphology, as shown in Fig. S8. The HDFs were prestained with CellTracker Deep Red. After 24 hours, the cells are fixed and stained with DAPI (nuclei, blue) and Phalloidin (actin, green), as shown in the first column in Fig. \ref{fig:patterned_defects}. We used a method similar to that described in the previous section to visualize the density map, as shown in the second column in Fig. \ref{fig:patterned_defects}. Then we compute the local density of myofibroblasts $\phi_\mathrm{MF}$, by counting the number of fibroblasts and myofibroblasts, within every annulus at an increment of $\Delta r$ = 30 $\mu$m from the center of the defect $|\mathbf{r}-\mathbf{r}_\mathrm{defect}|$. For comparison, we approximate the myofibroblast concentration $\Phi_\mathrm{MF}$ of the sample (dashed lines) by calculating its number density over the entire field of view, shown as the dotted lines in the third column. We find that, similar to mobile defects, -$\frac{1}{2}$ defects serve as sites where myofibroblasts concentrate, whereas +$\frac{1}{2}$ defects exhibit an excess of fibroblasts. These trends persist over distances exceeding 200 $\mu$m, well beyond the defect core of the fiber pattern. The presence of isolated, pinned defects with distinct cell-type selectivity supports the observed dynamics of defect movement. We next postulate the mechanism underlying this relative enrichment compared to the bulk.

Defects, as sites of strong distortion and stress concentration, also correlate with rare events such as apoptosis \cite{saw2017topological} and proliferation \cite{endresen2021topological}. In particular, positively charged defects are associated with local compressive stresses that were found to associate with cell extrusion, apoptosis, and morphogenesis \cite{maroudas2021mechanical,endresen2021topological,guillamat2022integer}. Our previous work \cite{luo2025rigidity} demonstrates that fibroblasts and myofibroblasts differentially interpret stiffness heterogeneity, with myofibroblasts exhibiting greater sensitivity to stiffness gradients and generating stronger contractile forces in response to cell-scale features. We hypothesize that different phenotypes can also exhibit differentiated affinity for distinct mechanical microenvironments.  If so, we would also expect myofibroblasts to be enriched at negatively charged defects and depleted at positively charged defects of higher charge. Therefore, we examine whether integer defects ($\pm1$) also preferentially colocalize with specific cell types. Since $\pm 1$ defects spontaneously split into lower charges in 2D, we do not observe them on featureless substrates. Nonetheless, we may pattern integer defects using LCE fiber substrates (Fig. \ref{fig:patterned_defects}c,d) in the same manner, as defects of integer charges can sometimes appear at physiological or pathological settings, such as the tumor-stromal interface \cite{provenzano2006collagen}, or the optical nerve ending \cite{miranda2022radial}. 

We observe a clear excess of myofibroblasts near the core of -1 defect (Fig. \ref{fig:patterned_defects}c), and a mild excess of fibroblasts near the radial +1 defect (Fig. \ref{fig:patterned_defects}d), consistent with $\pm\frac{1}{2}$ defects. Potential factors obscuring the signals may stem from the splitting of higher-charged defects within the mixed-cell monolayer due to finite anchoring at the cell-LCE interface, which have been discussed in Ref. \cite{turiv2020topology}. While the precise split distance of the mixed phenotypes is unknown, it can be approximated using the value for a pure HDF layer, where an integer defect typically separates into pairs spaced 200 to 500 $\mu$m apart \cite{turiv2020topology}. Instead of a single defect, two nearby defects may each modulate the local density while simultaneously interacting with each other. Nonetheless, when we repeat this experiment with slightly different initial $\Phi_\mathrm{MF}$, we confirm that myofibroblasts are enriched in negatively charged defects and fibroblasts are enriched in positively charged defects in several other cases (Fig. S10).

Why do myofibroblasts favor negatively charged defects, while avoiding positively charged ones? The clue may lie in the type of stress near defects, modulated by factors including the rate of cell division \cite{kaiyrbekov2023migration}. An important mediator of nuclear responses is Yes-associated protein (YAP), an apoptosis inhibitor and downstream transducer of the Hippo signaling cascade. YAP functions as a cellular barometer for sensing local crowding \cite{panciera2017mechanobiology}. YAP preferentially localizes to the nuclei (i.e., becomes activated) at sites of negatively charged defects \cite{endresen2021topological,saw2017topological} and becomes deactivated at high cell densities \cite{chen2024disodium} through a process known as contact-inhibited proliferation \cite{pavel2018contact}. Consistent with these observations, we find that myofibroblasts exhibit elevated YAP activity near the core of $-\frac{1}{2}$ defects (Fig. S11). While myofibroblasts tend to have more activated YAP even when they are isolated \cite{luo2025rigidity}, recent work indicates that, in the cell collective, density-dependent responses override substrate stiffness and other single-cell–level attributes in governing YAP activation \cite{grudtsyna2025packing}.

These observations suggest two non-exclusive possibilities: The intrinsically high YAP activation in myofibroblasts promotes local contractile force patterns that favor the formation of $-\frac{1}{2}$ defects, or conversely, myofibroblasts preferentially migrate to sites where the local stress landscape promotes YAP activation. On the one hand, characteristic patterns of motion, traction, and stress appeared long before defect formation \cite{bera2025traction}. This suggests that defects originate from pre-existing, coordinated cell-force and motility patterns, consistent with our observation that incorporating myofibroblasts slows down the defect recombination. On the other hand, our previous study \cite{chen2024disodium} showed that collectively, cells can generate extensile internal stress to offset the effect of contact-inhibited proliferation, leading to reduced deactivation of YAP.  The net result appears to be that myofibroblasts preferentially occupy negatively charged defects by avoiding regions with compressive stress (+1 and $+\frac{1}{2}$ defects), which can lead to apoptosis. More myofibroblasts lead to more persistent defects and vice versa, establishing a feedback loop.

\section{Discussion}

    While extensive studies have characterized the steady-state properties of cellular assemblies of various degrees of order, time-resolved, large-scale studies of defects and their role in governing collective dynamics remain scarce. 
In this study, we aim to derive a self-organization rule governing the myofibroblast–fibroblast mixture, where one phenotypes migrate more slowly than the other. 
We resort to understanding this phenotype-specific restructuring in 2D monolayers at high density, where collective dynamics emerge. Unlike previous studies on the biochemical and phenotypic responses of individual cells, we seek to elucidate how composition influences the emergent behavior, through large-scale live cell microscopy of monolayers with an increasing fraction of myofibroblasts, mimicking fibrosis progression in vitro.

 A multitude of studies \cite{saw2017topological,sarkar2023crisscross,kawaguchi2017topological} report the biological functions of defects, which are topologically and mechanically distinct from the surrounding tissues. Overall, in this system, we observe the emergence of robust nematic order in the broad range of myofibroblast concentration studied. The dynamics are primarily governed by the spontaneous recombination of $\pm\frac{1}{2}$ defects. By using an analytical method that conserves topological defect charges, we track movements and densities of defects over time. We observe that higher $\Phi_\mathrm{MF}$ leads to reduced defect recombination rates, while the steady-state defect density remains largely unaffected. The former indicates that the slowed defect dynamics are likely due to the downregulated motility of myofibroblasts compared to fibroblasts. The latter suggests that, over long timescales, equilibrium defect densities reflect similar overall activity across different systems. Despite the reduction in the magnitude of the velocity of both the defect and the cells with increasing $\Phi_\mathrm{MF}$, we find that the spatial velocity correlations remain similarly long-ranged, over hundreds of $\mu$m, across all $\Phi_\mathrm{MF}$. These observations echo recent theoretical studies of quenched nematics \cite{kumar2020active,kumar2022active,kinoshita2025active} where noise is frozen in space. Our observations suggest that adding a slower-moving phenotype can produce some of the same effects.

We show that the slowed dynamics of defect recombination hinder the myofibroblasts from achieving well-ordered alignment under externally imposed, sparsely-spaced, unidirectional alignment cues. The reduced ability to maintain uniform alignment is reflected in a more spread-out distribution of angle deviations from the preferred direction. We thereby establish means to examine the collective behavior of a dense monolayer, providing a straightforward way to quantitatively determine the fraction of myofibroblasts.

Depending on their signs, $\pm \frac{1}{2}$ defects often exhibit distinct characteristics due to their different symmetries.  Positively-charged, $+\frac{1}{2}$ defects are highly mobile \cite{balasubramaniam2021investigating,kawaguchi2017topological} and primarily responsible for driving recombination dynamics. They also tend to be confined within regions of high activity \cite{zhang2021spatiotemporal,chaithanya2024transport}. Similarly, while $-\frac{1}{2}$ defects can be easily immobilized by the pillars, $+\frac{1}{2}$ defects remain motile in an active nematic composed of motile bacteria \cite{figueroa2022non}. It is known that cells with distinct contractile and extensile forces self-sort based on the nature of their forces \cite{balasubramaniam2021investigating,sahu2020small}. In this work, we find that cells with distinct forces also respond differently to different mechanical stresses at topological defects. Our analysis reveals that myofibroblasts have an affinity to $-\frac{1}{2}$ defects while the fibroblasts tend to localize at $+\frac{1}{2}$ defects.  
The depletion of myofibroblasts from the more active  $+\frac{1}{2}$ defect sites is also consistent with the slowdown of the overall recombination dynamics of the monolayer as  $\Phi_\mathrm{MF}$ increases.

Open questions persist regarding the precise mechanobiological transduction mechanisms at defect sites. 
A central debate is whether defects are drivers of biological processes \cite{saw2017topological,guillamat2022integer} or merely symptoms of the underlying heterogeneity, driven by spatial coordination of cell movements \cite{bera2025traction}, cell shape deformation and force transmission \cite{mueller2019emergence}, orientational or hydrodynamic fluctuations \cite{bonn2022fluctuation}.
 We find that this phenotype-selective mechanism at defect cores applies even to preexisting pinned defects and defects of higher charges, supporting the idea that cells respond to distinct force environments at defects. Our results also indicate that mechanical forces associated with topological defects can influence downstream protein expressions through an outside-in process. We find that, consistent with previous studies, the apoptosis inhibitor YAP is deactivated near positively charged defects \cite{saw2017topological} but remains active near negatively charged defects \cite{endresen2021topological}. These effects are emergent and arise collectively from the interaction between the phenotypes. 

We demonstrate that myofibroblast preferentially colocalized with defects of negative charges, causing increased friction that slows global defect recombination.
The observed behavior is reminiscent of recent theoretical works on active nematics with quenched disordered \cite{kumar2020active,kumar2022active,kinoshita2025active}, where increasing disorder strength slows down defect recombination. We hope this work motivates further experiments and theoretical studies on the interplay between cell phenotype and topological defect.

\section{Materials and Methods}
\subsection{Chemicals}
All chemicals were purchased from Sigma-Aldrich (St. Louis, MO), and used without modification unless otherwise stated.
Azodye SD1 dye was purchased from DIC Corporation (Tokyo, Japan). Reactive monomer RM257 was purchased from Ambeed (Arlington Heights, IL), and a mixture of small molecule liquid crystal mixture was purchased from Instec Inc (Boulder, CO, catalog no. LC-VATS14). 
Dulbecco’s Modified Eagle's Medium (DMEM), DMEM F12 Medium, fetal bovine serum (FBS), phosphate-buffered saline (PBS), Bovine Serum Albumin (BSA), 100x antibiotic-antimycotic, and 100x penicillin-streptomycin (pen-strep) were purchased from Gibco (Grand Island, NY). 
Transforming growth factor-beta 1 (TGF-$\beta$1, ab50036), FITC anti-alpha smooth muscle actin antibody (ab8211), and phalloidin-iFluor 555 (ab176756) were purchased from Abcam (Cambridge, United Kingdom). Hoechst 33342, 16 wt\% paraformaldehyde solution (PFA), CellTracker in Deep Red (C34565) and Green (C7025) were purchased from  ThermoFisher (Waltham, MA).  Primary anti-rabbit YAP (NB110-58358) was purchased from Novus Biologicals (Centennial, CO). The secondary antibody Alexa Fluor 488 AffiniPure Goat Anti-Rabbit IgG (AB2338052) was purchased from Jackson ImmunoResearch Inc (West Grove, PA). 
Water was dispensed from a Milli-Q system with resistivity 
 = 18.2 M$\Omega\cdot$cm at 25$^\circ$C.

\subsection{Cell acquisition, culturing, and treatments} 

Adult human dermal fibroblast (HDFa) cells were purchased from the American Type Culture Collection (ATCC, Manassas, VA, catalog no. PCS-201-012). 
HDFs were maintained at 37$^\circ$C with 5\% CO$_2$ in complete media composed of DMEM, supplemented with 10 vol\% FBS, and 1x pen-strep. The media was refreshed every 2 days, and the cells were sub-cultured upon reaching 70 - 80\% confluency. Fibroblasts were labeled by CellTracker Deep Red. Early passages (P$<$5) were used for induction. To induce the myofibroblast phenotype in vitro, we first cultured HDFa to confluency, and then grew them in serum-starved media for 24 hours. Cells were rinsed with PBS, and incubated in serum-free media with 10 ng/ml TGF-$\beta$1 for 96 hours. After 48 hours, cell media was refreshed, and fresh  TGF-$\beta$1 was added. Myofibroblasts were labeled by CellTracker Green.

\subsection{Live cell microscopy}

The 4-well glass-bottom Petri dishes (35 mm, Cellvis, Mountain View, CA, catalog D35C4-20-0-N, \#1.5 cover glass) were pre-coated with a low-concentration collagen solution and rinsed 3x with PBS. 
Cells were seeded at $\rho_\mathrm{cell} \approx$ 750 mm$^{-2}$ density. Live cell imaging began as soon as the cells had attached to the bottom of the dish (approximately 6 hours after seeding). Cells were maintained in onstage incubators at 37 $^\circ$C under 5\% CO$_2$, and imaged using a 10x air objective (NA = 0.8), 4x4 binning, acquired in both far red (peak wavelength = 602 nm) and green (peak wavelength = 517 nm), at a time interval of every 30 minutes. The pixel size is 2.344 $\mu$m. Approximately 100 frames were acquired for each $\Phi_\mathrm{MF}$. Tens of images were stitched together following procedures in \cite{luo2022cell} using an automated stage and the ZEN software (Carl Zeiss AG, Germany).

\subsection{Immonoflurescence}
Prior to observation, cells were fixed using a 4 wt\% PFA, and stained with antibodies and fluorescent markers.
Actin and nuclei were stained by Phalloidin-iFluor 555 (1000$\times$ dilution) and Hoechst 33342 (1 $\mu$g/mL) following standard protocols \cite{luo2022cell}. YAP staining was performed as described in Refs. \cite{das2016yap,chen2024disodium}.

\subsection{Defect identification}
Briefly, raw phase contrast or fluorescence microscopy images were preprocessed to enhance cellular features. Cell orientation fields were calculated using the structure tensor method, by applying a Gaussian smoothing ($\sigma$ = 9.4 $\mu$m or 4 pixels) filter to 32 pixel $\times$ 32 pixel windows (75 $\mu$m $\times$ 75 $\mu$m). Defect detection was implemented by adopting the algorithm from \cite{tan2019topological} by identifying patch regions of grid points undergoing rapid rotation ($\Delta \theta > \frac{\pi}{4}$) of the director field. The topological charge for each patch was computed from   ($s = \frac{\oint \Delta\theta}{2\pi}$), summed along the path of the loop. This way, topological charges are conserved. Further details can be found in the Supplementary Information (SI).

\subsection{Fabrication of LCE fiber substrates with different topological charges}

Glass slides (AmScope, Irvine, CA) were thoroughly cleaned with detergent in an ultrasonic bath for 8 minutes and then rinsed with water, acetone, and isopropanol. After drying in the oven at 80$^\circ$C for 10 minutes to evaporate solvents, they were further cleaned with a plasma cleaner (Harrick Plasma, Ithaca, NY) for 5 minutes. The fibers were prepared by a projection display based on procedures detailed in \cite{chen2022nematic}. Briefly, a solution of azodye SD1 dye at 0.2 wt \% in N,N-dimethylformamide (DMF) was evenly deposited on the glass slide by spin coating at 3000 rpm for 30 sec, and annealed at 120 $^\circ$C for 15 min. Two such coated glass slides were assembled into a chamber with coated sides facing each other, separated by spacer beads (Cospheric, Somis, CA, diameter 2$a$ = {20} $\mu$m).  Then, patterns were generated using a rotating linear polarizer in front of the projection display. A solution of 7 wt\% reactive monomer RM257, 92 wt\% LC mixture, and 1 wt\% photoinitiator was introduced into the chamber at isotropic temperature (120$^\circ$C) and slowly cooled down to the nematic phase. The sample was cured under UV light with power 6 mW/cm$^2$ at 365 nm wavelength for 1 h to crosslink the reactive monomer completely. Afterward, the samples were immersed in hexane overnight to wash away the uncrosslinked small-molecule LC. Finally, the samples were dipped in liquid nitrogen for a few seconds and split with a razor blade.

\section*{Acknowledgements}
The authors thank Toshi Parmar, Jacob Notbohm, and Cristina Marchetti for stimulating discussions about active nematics. We acknowledge Kyogo Kawaguchi and Kevin Mitchell for sharing the codes for determining director-field orientation and for identifying topological defects, Shervin Issakhani for initial code development, Valerie Horsley and Elizabeth Caves for providing the healthy and SSc cells from donors, and for the helpful discussion on fibrotic markers and media conditions, and Nick Bernardo for assistance with machining the imaging stage.

\section*{Funding}
This work is partially supported by the National Science Foundation under the Grant OAC-2411044.

\section*{Author contributions}
Yimin L designed the project. Yuxin L, JC, and ZZ performed experiments. ZZ and Yimin L performed numerical analysis. All authors have contributed to the writing of the manuscript and approved the final version of the manuscript.

\section*{Conflicts of interest}
The authors declare no conflicts of interest. 

\section*{Data availability statement}
The data collected and the analysis codes generated for this study will be made available through a permanent GitHub repository upon acceptance of the manuscript.

\clearpage
\setcounter{section}{0}
\setcounter{figure}{0}
\setcounter{table}{0}
\setcounter{equation}{0}
\renewcommand{\thefigure}{S\arabic{figure}}
\renewcommand{\thetable}{S\arabic{table}}
\renewcommand{\thesection}{Section S\arabic{section}.}
\renewcommand{\theequation}{S\arabic{equation}}
\captionsetup[figure]{labelfont={bf},labelformat={default},labelsep=period,name={Figure}}

\begin{center}
{\LARGE \textbf{Supplementary Information}}\\[0.5em]
{\large Myofibroblasts slow down defect recombination dynamics in mixed cell monolayers}
\end{center}

\textbf{The PDF file includes:}\\
Additional experimental and analysis descriptions\\
Supplementary Figure S1 to S11\\
Supplementary Table S1 to S2 \\

\textbf{Supplementary video S1.} Live-cell microscopy of a 50:50 mixture of HDF (magenta) and MF (green) at 750 cells / mm$^2$, play speed = 30 frames / sec, real speed = 30 minutes / frame, speed up = 18000 $\times$ real time. The monolayer exhibited correlated motion at this concentration.

\textbf{Supplementary video S2.} Live-cell microscopy of a 50:50 mixture of HDF (magenta) and MF (green) at 500 cells / mm$^2$, play speed = 30 frames / sec, real speed = 30 minutes / frame, speed up = 18000 $\times$ real time. The monolayer exhibited little coherent motion at this concentration.

\clearpage
\section{Healthy and SSc fibroblasts from donors}

\begin{figure}[htb!]
\begin{center}
    \includegraphics[width=0.7\textwidth]{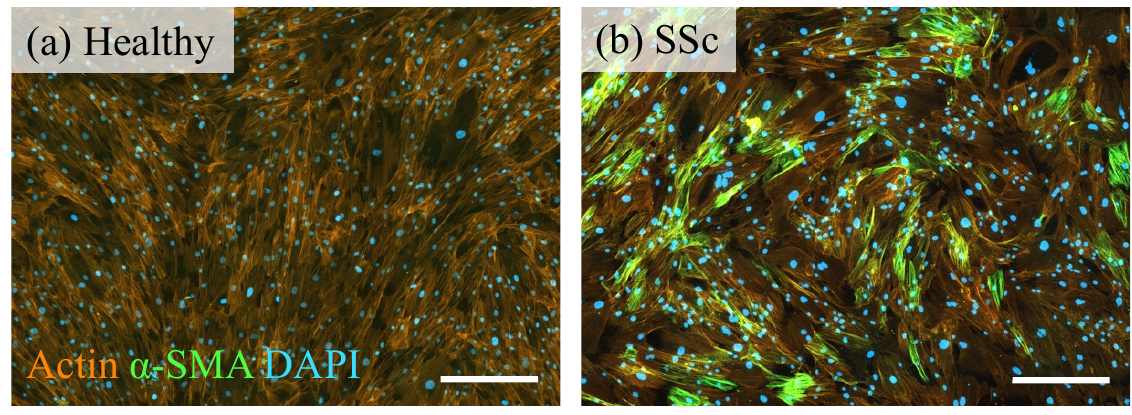}
\end{center}
  \caption{Immunofluorescence staining of dermal fibroblasts from (a) healthy and (b) SSc donors, showing actin (orange), $\alpha$-SMA (green), and nuclei (blue). The SSc cells overexpress $\alpha$-SMA. The scale bars are 200 $\mu$m.}
\label{fig:health_vs_diseased}
\end{figure}

Healthy and SSc fibroblasts were generous gifts from the Horsley Lab at Yale, and were isolated from discarded biopsy samples following the protocol outlined in Ref. \cite{vangipuram2013skin}, and have been obtained under informed consent and conform to HIPAA regulations. These donor-sourced fibroblasts were maintained and passaged in media consisting of DMEM F12 medium,  supplemented with 10 vol\% FBS, and 1x antibiotic-antimycotic, and incubated at 37$^\circ$C with 5\% CO$_2$. To visualize $\alpha$-SMA (Fig. \ref{fig:health_vs_diseased}), the cells are incubated in FITC anti-alpha smooth muscle actin antibody (200x dilution) with PBS overnight at 4$^\circ$C. SSc fibroblasts have an overexpression of $\alpha$-SMA in the actin (Fig. \ref{fig:health_vs_diseased}b), compared to cells from healthy donors (Fig. \ref{fig:health_vs_diseased}a). Immunofluorescence analysis shows that approximately 50\% of the cell population is $\alpha$-SMA-positive, and the cell orientation exhibits local nematic order.

\clearpage
\section{Maintaining myofibroblasts in vitro}

\begin{figure}[htb!]
\begin{center}
    \includegraphics[width=0.65\textwidth]{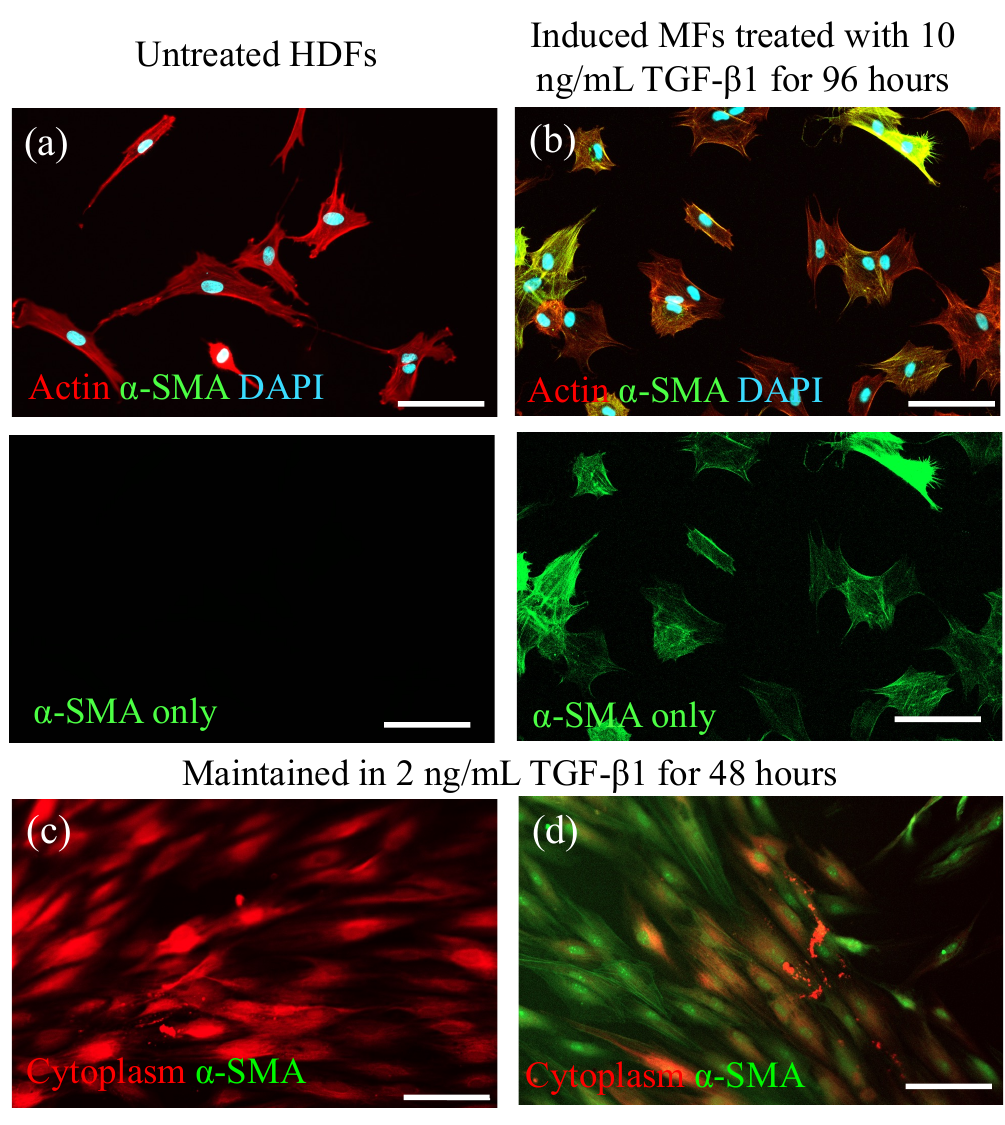}
\end{center}
  \caption{Immunofluorescence micrographs of the two phenotypes. The fibroblasts have elongated morphology, whereas myofibroblasts are more rounded and express $\alpha$-SMA (green) in their actin (magenta). (a) Untreated human dermal fibroblast cells. (b) Cells induced with 10 ng/mL TGF-$\beta$1 for 96 hours overexpress $\alpha$-SMA. 
 (c) Untreated fibroblasts maintain their unactivated phenotype after exposure to 2 ng/mL TGF-$\beta$1 for 48 hours. (d) On the contrary, after induction, the majority of myofibroblasts overexpress $\alpha$-SMA when maintained in 2 ng/mL TGF-$\beta$1 for 48 hours. The scale bars are 100 $\mu$m.}
\label{fig:phenotype_maintenance} 
\end{figure}

We induce myofibroblasts in vitro (Fig. \ref{fig:phenotype_maintenance}) using adult human dermal fibroblasts (HDF), by adding 10 ng/mL transforming growth factor-$\beta$1 (TGF-$\beta$1) into the media. The induced myofibroblasts exhibit an altered phenotype.  In isolation, HDFs exhibit elongated morphologies (Fig. \ref{fig:phenotype_maintenance}a), whereas myofibroblasts are more rounded (Fig. \ref{fig:phenotype_maintenance}b). 

The myofibroblast phenotype was maintained using a lower concentration of TGF-$\beta$1 (2 ng/mL). We also confirmed that under this reduced TGF-$\beta$1 concentration, fibroblasts retained their elongated morphology after 48 hours (Fig. \ref{fig:phenotype_maintenance}c), whereas myofibroblasts maintained their activated phenotype, characterized by $\alpha$-SMA expression (Fig. \ref{fig:phenotype_maintenance}d).

\clearpage
\section{Determining myofibroblast fraction $\mathrm{\Phi_{MF}}$}

Myofibroblast fractions are determined using three different methods for the same image, for two total cell concentrations, $\rho_\mathrm{cell}$ = 500 cells mm$^{-2}$ and $\rho_\mathrm{cell}$ = 750 cells mm$^{-2}$.

\textit{1) Segmentation by CellPose}: The myofibroblasts (green) and fibroblasts (magenta) channels were 
segmented using Cellpose \cite{stringer2021cellpose} to extract the position of each phenotype.

\textit{2) Cell ratio by area}: We estimate the areal fraction taken up by each cell phenotype by first binarizing the image to obtain the area of fibroblasts ($\mathrm{A_{magenta}}$) and myofibroblasts ($\mathrm{A_{green}}$), and computing the areal fraction: $\mathrm{\Phi_{MF}=\frac{A_{green}}{A_{magenta}+A_{green}}}$. 

\textit{3) Nuclei by TrackMate:} Nuclei are first identified using TrackMate and overlaid with the fibroblasts (magenta) channel to identify those associated with the fibroblasts, and the rest of the nuclei are classified as myofibroblasts. 

\begin{figure}[htb!]
\begin{center}
    \includegraphics[width=.9\textwidth]{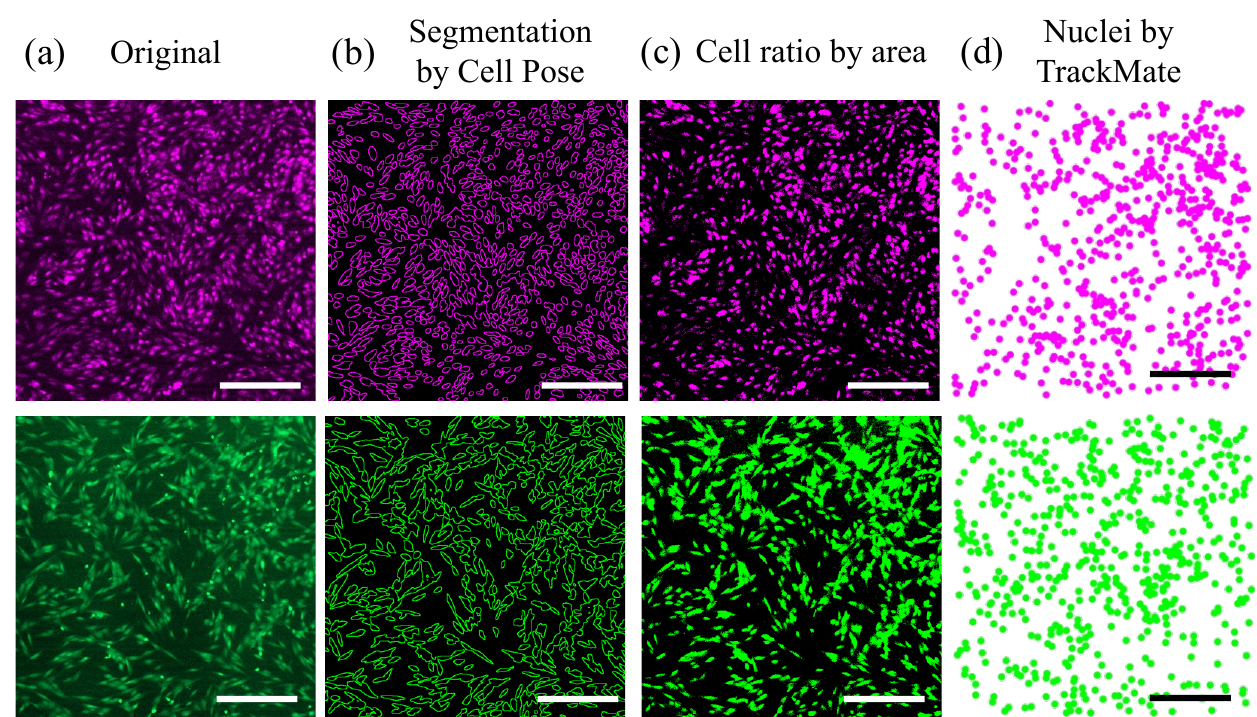}
\end{center}
  \caption{Example results for extracting cell concentrations of each type: (a) original image, (b) CellPose segmentation, (c) cell-type ratio obtained by binarization and area quantification, and (d) assignment of each nucleus to either fibroblasts or myofibroblasts using TrackMate. The scale bars are 500 $\mu$m. }
\label{fig:MF_fraction}
\end{figure}

Measured myofibroblast fractions using all methods are close to 50\% and align with the experimental design. At very high cell densities, myofibroblasts become compressed by surrounding cells, causing them to occupy a 2D cross-sectional area similar to that of fibroblasts—even though they appear larger when isolated.
     
\begin{table}[ht]
\centering
\caption{Determination of myofibroblast fraction using different methods}
\label{table:count_comparison}
\begin{tabular}{lcc}
\toprule
 & $\rho_{\text{cell}} = 500\ \text{cells mm}^{-2}$ & $\rho_{\text{cell}} = 750\ \text{cells mm}^{-2}$ \\
\midrule
Experiment design  & 0.50 & 0.50 \\
1) Segmentation by CellPose  & 0.45 & 0.44\\
2) Cell ratio by area  & 0.47 & 0.44\\
3) Nuclei by TrackMate   & 0.49 & --\\
\bottomrule
\end{tabular}
\end{table}

\newpage
\section{HDF and MF velocity}
At a lower density of 500 cells/mm², cells are better separated, and experience fewer interactions with neighboring cells, allowing their speed to be tracked more accurately. We track nuclei trajectories while also assigning the cell types by overlapping with either magenta (HDF) or green (MF) channels. We find that fibroblasts ($\bar{v}_\mathrm{HDF}$ = 6.97 $\pm$ 0.02 $\mu$m/h) move faster than MF ($\bar{v}_\mathrm{MF}$ = 5.73 $\pm$ 0.02 $\mu$m/h). The analysis includes 98,890 HDF and 71,832 MF tracked across 100 frames of experimental images. One-way ANOVA finds that the difference is highly significant. 

\begin{figure}[htp!]
    \centering
    \includegraphics[width=\linewidth]{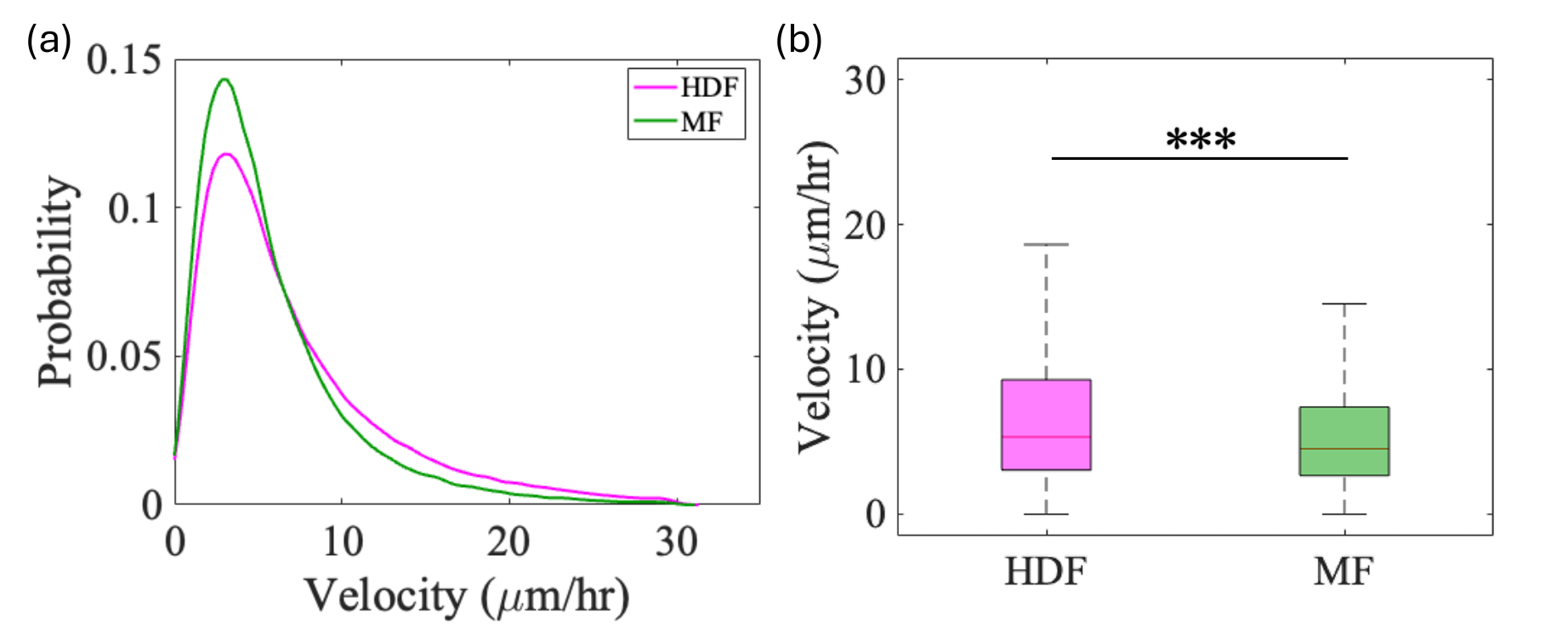}
    \caption{Velocity comparison between HDF and MF. (a) Distribution of velocities for HDF and MF. (b) Box plot of HDF and MF velocity, one-way ANOVA test found p $<$ 0.0001.}
    \label{fig:placeholder}
\end{figure}

\clearpage

\section{Particle Image Velocimetry (PIV) and cell velocity correlation procedures}

Velocity flow within the monolayer is quantified by using the PIVlab toolbox in MATLAB. For each condition, we select two 1875 $\mu$m $\times$ 1875  $\mu$m, and two frames separated by 2.5 hours, to compute the displacement by PIV.  In the pre-processing step, we use the Contrast Limited Adaptive Histogram Equalization (CLAHE) technique with a window size of 64 pixels (= 150 $\mu$m) to enhance local contrast and minimize illumination inhomogeneity.
Then we apply a multipass fast Fourier transform window deformation with a pass interrogation length of 60 pixels (= 140 $\mu$m) and a step size of 30 pixels  (= 70 $\mu$m) to improve the accuracy and resolution of flow velocity measurements. Sub-pixel displacement estimation was performed by fitting a 2D Gaussian function to the cross-correlation peak, to enable interpolation of the displacement vector position with sub-pixel accuracy. Vector validation was implemented to remove spurious measurements that may arise from low image correlation or poor signal-to-noise ratio. Specifically, a universal outlier detection filter was applied, where velocity vectors deviating by more than eight standard deviations from the local mean were removed.

\begin{figure}[htb!]
\begin{center}
    \includegraphics[width=\textwidth]{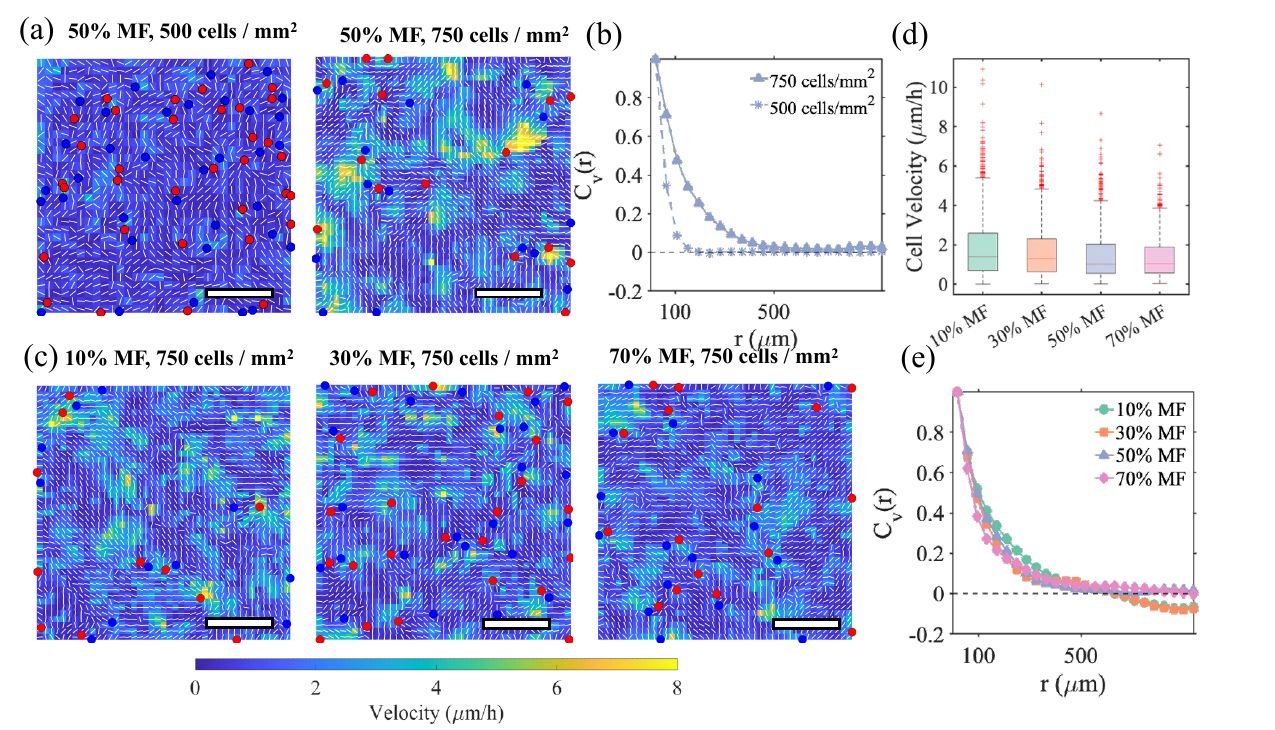}
\end{center}
  \caption{(a, c) Representative velocity-magnitude maps overlaid with director fields (light-grey segments), for monolayers with cell densities of $\rho_\mathrm{cell} = $ 500 mm$^{-2}$ (left) and $\rho_\mathrm{cell} = $ 750 mm$^{-2}$ (right), at 50\% myofibroblasts. (b) The spatial velocity correlation function for different densities. (c)  Representative velocity-magnitude maps overlaid with director fields for varying myofibroblast concentrations at the same density of $\rho_\mathrm{cell} = $  750 mm$^{-2}$. (d) Box plots and (e) The spatial velocity correlation function for different MF concentrations. }
\label{fig:piv_diff_MF}
\end{figure}

These results are shown in Fig. \ref{fig:piv_diff_MF}a,c. For analysis to compute the spatial velocity correlation, modified from \cite{kinoshita2025active}. 
\begin{equation}
    C_{\mathrm{v}}(\mathbf{r}) = \frac{\langle\mathbf{v}(\mathbf{r}')\cdot \mathbf{v}(\mathbf{r}'+\mathbf{r})\rangle_{\mathbf{r}', |\mathbf{r}| = r}}{\langle\mathbf{v}(\mathbf{r}')\cdot \mathbf{v}(\mathbf{r}')\rangle}.
\end{equation}

At 50\% myofibroblast concentration and $\rho_\mathrm{cell}$ = 500 mm$^{-2}$, we observe no correlated motion (Fig. \ref{fig:piv_diff_MF}b) compare to monolayers at the same myofibroblast fraction and $\rho_\mathrm{cell}$ = 750 mm$^{-2}$, cell-cell correlation falls on the order of a single cell ($\sim$100 $\mu$m, Fig. \ref{fig:piv_diff_MF}b). Compared to defect velocities ($\sim$ 10s $\mu$m/h), the cell velocities are much slower ($\sim$ $\mu$m/h). This is not surprising as topological defects routinely move much faster than the underlying materials, because defect motion is governed by energetic and structural considerations, not by the bulk transport of matter. We also find that, despite the variation in average cell velocity (Fig. \ref{fig:piv_diff_MF}d), the spatial correlation of the velocity field does not show a clear dependence on myofibroblast concentration (Fig. \ref{fig:piv_diff_MF}e). 


\clearpage

\section{Director field and defect identification by topological methods}

The director field is calculated by using the tensor method of calculation:
$G_i(i =x,y)$ is the intensity gradient in the $i$-th direction ($i$ = 1, 2).
\begin{equation}
A_{xx} = G_x^2, \quad A_{yy} = G_y^2, \quad A_{xy} =A_{yx} =  G_x G_y,
\end{equation}

These terms make up the gradient tensor:
\begin{equation}
    \Gamma = \begin{bmatrix}
A_{xx} & A_{xy} \\
A_{yx} & A_{yy}
\end{bmatrix},
\end{equation}

\begin{figure}[htb!]
\begin{center}
    \includegraphics[width=1\textwidth]{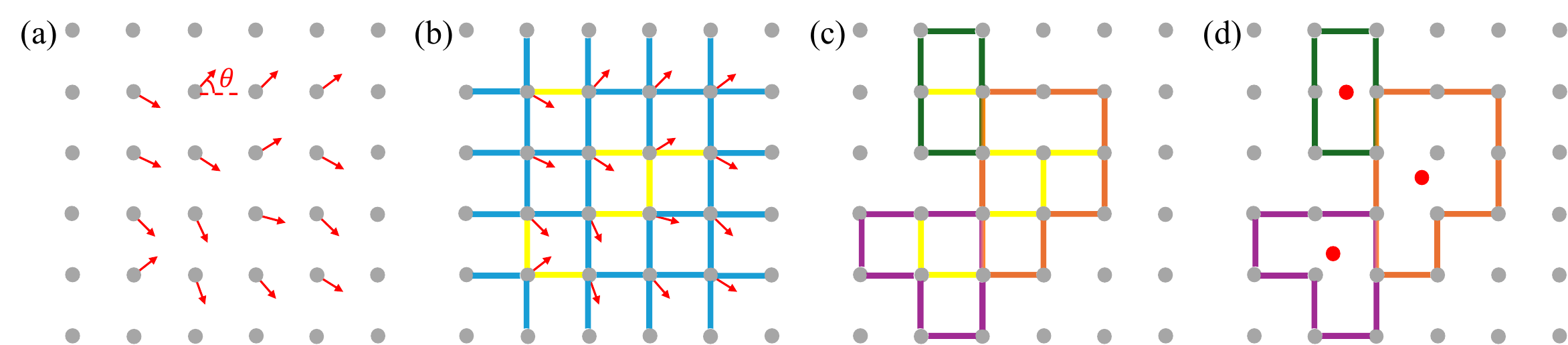}
\end{center}
  \caption{Schematic illustrating the procedure for identifying defect positions. (a) Gray dots represent the grid points, and red arrows indicate the local orientation angle $\theta$. (b) Yellow segments denote regions with sharp angular changes, while blue segments correspond to smooth angular variations between neighboring grid points. (c) Green, purple, and orange curves mark the smallest closed loops enclosing clusters of connected yellow segments. (d) Solid dots indicate the centers of these loops, which are assigned as the defect positions. The topological charge of each defect is determined from the winding number. Red dots, representing +$\frac{1}{2}$ defects, are shown here for illustrative purposes.}
\label{fig:rotation_director}
\end{figure}

The orientation angle $\theta$ is the local direction of the director field with respect to the $x$-axis, which is given by
\begin{equation}
\theta = \frac{1}{2} \arctan\left( \frac{2 A_{xy}}{A_{xx} - A_{yy}} \right) + \frac{\pi}{2}.
\end{equation}
They are shown in Fig. \ref{fig:rotation_director}a as red arrows.

To preserve the topology of defects, we identify the defects using the method described in Ref. \cite{tan2019topological}. A closed loop within a region is identified as undergoing a significant change in orientation if the angle change $\Delta\theta$between neighboring grid points exceeds a threshold of $\frac{\pi}{4}$. Lines connecting neighboring grid points will be categorized into two types: smooth angle change ($\Delta\theta<\frac{\pi}{4}$, shown as blue segments) and sharp angle change ($\Delta\theta\ge\frac{\pi}{4}$, shown as yellow segments). Segments exhibiting large angular changes indicate the presence of nearby defects. Therefore, the smallest closed loops are selected to encompass all connected yellow segments (Fig. \ref{fig:rotation_director}c). Winding number is computed by summing up the total rotation as one goes around the closed loop, to compute the topological charge of the enclosed defect, $\oint \Delta\theta$, that can only be $\pm\pi$, resulting in defects of charge of either $+\frac{1}{2}$ or $-\frac{1}{2}$, given that integer charge defects are rarely observed in 2D due to their high energetic cost. For simplicity, the center of each loop is set to be the defect position (Fig. \ref{fig:rotation_director}d).

\clearpage
\section{Defect Density evolution at all myofibroblast concentrations}

Initial defect density $\rho_0$ is highly influenced by initial cell seeding conditions. When only the bottom surface of the imaging dish is collagen-treated, cells lack any preferred orientation upon attachment, leading to significant variability in defect density across cell concentration ratios and even within individual samples. Consequently, $\rho_0$ does not exhibit a clear dependence on cell concentration. With defect recombination, defect density is expected to plateau over time. However, experimental measurements show no trend in the long-time value of cell density $\rho_{\infty}$ across different samples and across different myofibroblast concentrations. Furthermore, $\rho_\infty$  does not reach zero, which can be attributed to factors such as cell activity (e.g. from either cell division or motility), local dewetting of the cell layer, and variability in the defect detection algorithm. However, the characteristic decay time constant, which describes the transition from $\rho_0$  and $\rho_\infty$, appears to follow a predictive trend with different myofibroblast concentrations. Each cell experiment is analyzed using three cropped regions of 800 pixels $\times$ 800 pixels (1872 $\mu$m $\times$ 1872 $\mu$m). Each region was analyzed individually to find the concentration of myofibroblasts. The time evolution of defect density for all samples is presented below in Fig. \ref{fig:density_decay_diff_conc}.

\begin{figure}[htb!]
\begin{center}
    \includegraphics[width=0.9\textwidth]{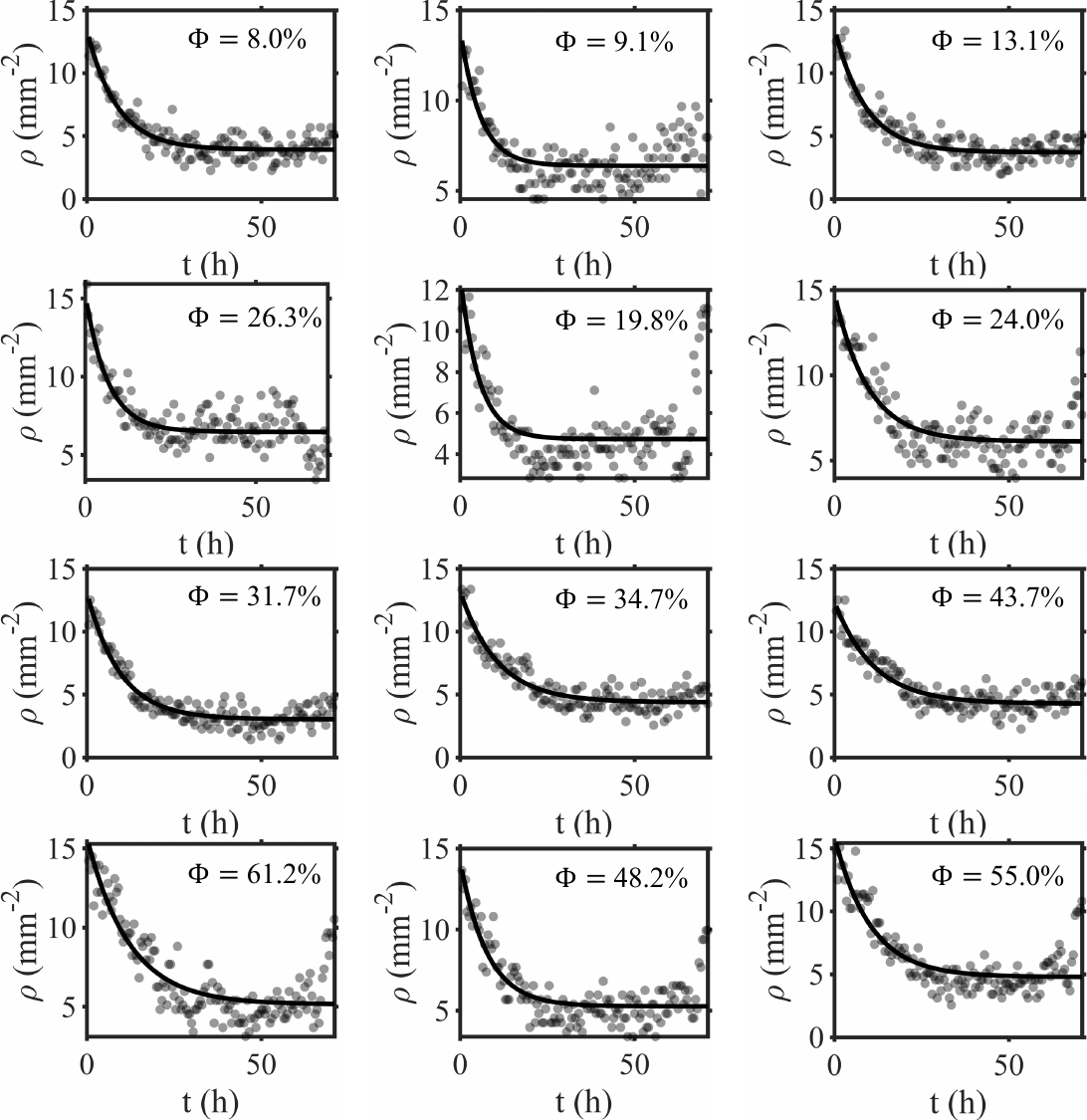}
\end{center}
  \caption{Density of $+\frac{1}{2}$ topological defects over time. The solid lines illustrate the exponential fit. Composition and fitting parameters are labeled in each sub-figure. }
\label{fig:density_decay_diff_conc}
\end{figure}

\newpage
\section{Scanning electron microscopy (SEM)}
\begin{figure}[htb!]
    \centering
    \includegraphics[width=1\linewidth]{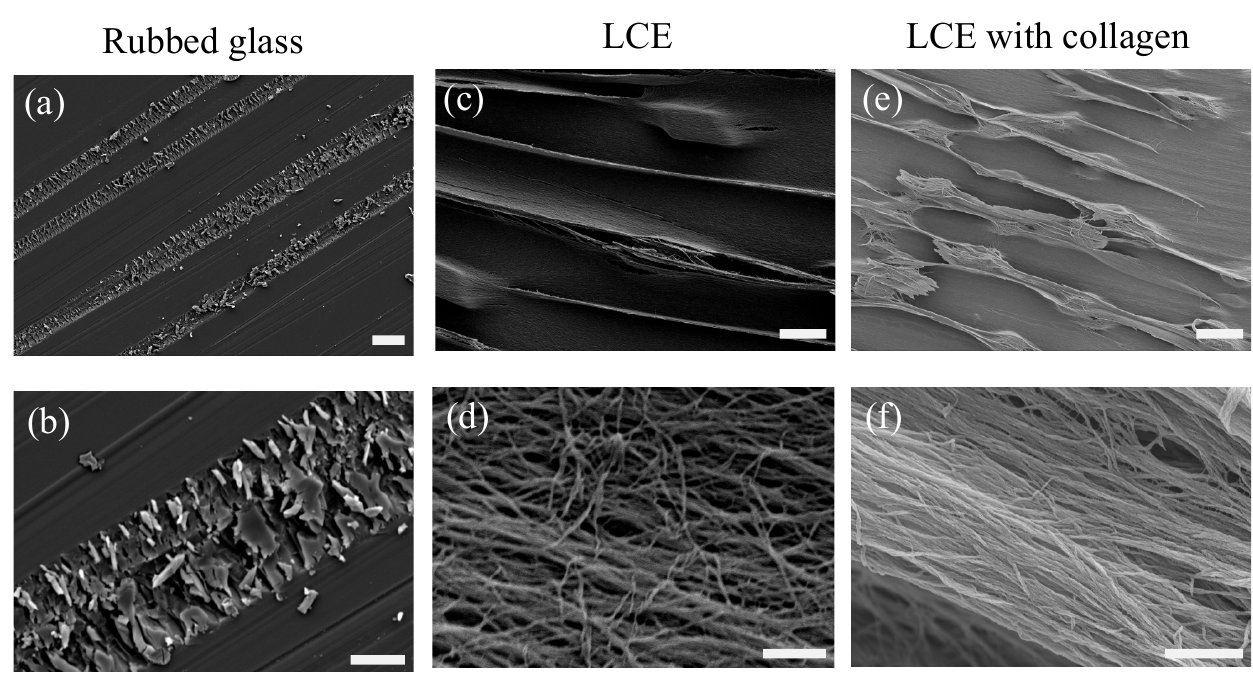}
    \caption{Scanning electron microscopy characterization of substrates. (a–b) The glass surface after rubbing with sandpaper. (c–d) Neat LCE fiber substrates without treatment. (e–f) LCE fiber substrate coated with collagen. Scale bars: (a) 80 $\mu$m; (c, e) 500 $\mu$m; (b, d, f) 20 $\mu$m. }
    \label{fig:SEM}
\end{figure}

In this work, both rubbed glass and LCE substrates were used to provide external alignment cues that influence cell alignment (Fig. \ref{fig:SEM}). The surface morphologies were characterized by a Hitachi SU-70 scanning electron microscope. All substrates were sputter-coated with a 4-nm iridium layer prior to imaging. Collagen coating does not change the surface morphology.

\clearpage

\section{Defect classification}

We first identify the position of fibroblasts and myofibroblasts. Then the local number density of fibroblast and myofibroblast, $n_\mathrm{HDF}$ and $n_\mathrm{MF}$ at each pixel is computed by binning over the surrounding 2.5 $\times$ 2.5 pixels (6 $\mu$m $\times$ 6 $\mu$m) area. The two channels were merged and overlaid with both the director field (black lines) and defects (see main text Fig. 4a).  We first manually categorize $\pm\frac{1}{2}$ defects into the MF/HDF category based on defect charge and the color of the cell density in their vicinity. Manually labeled images at various myofibroblast fractions are shown in the top row of Fig. 9.1. Defects of charges $+\frac{1}{2}$ and $-\frac{1}{2}$ are denoted by red and blue filled circles, respectively.
These results are tabulated in Table \ref{table:count_comparison}, columns 2-6.

\begin{figure}[htb!]
\begin{center}
    \includegraphics[width=\textwidth]{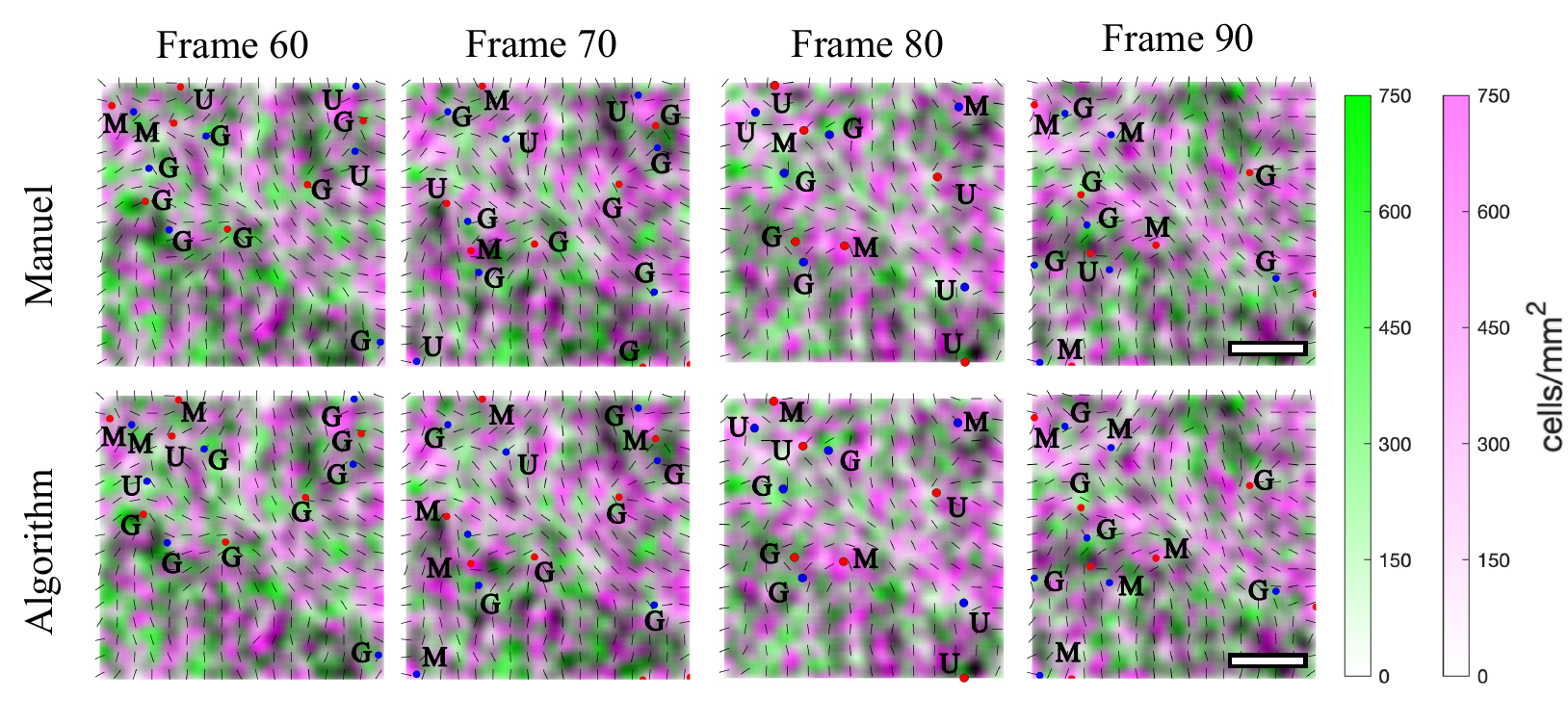}
\end{center}
  \caption*{\textbf{Figure S9.1.} Representative images comparing manual and automated labeling of myofibroblasts (green, ``G'') and fibroblasts (magenta, ``M''). Regions where the densities of these two cell types are comparable are marked as unclassified (``U''). }
\label{fig:density_determination}
\end{figure}

 \begin{table}[htb!]
 \begin{center}
     
\begin{tabular}{c|c|c|c|c|c|c|c|c|c|c}
 \toprule
 &  \multicolumn{5}{c|}{ \textbf{Manual count}}
  &  \multicolumn{5}{c}{ \textbf{Counting algorithm}}
  \\
   \cmidrule{2-11}
 \textbf{Snapshot ID} & MF  & MF  & HDF & HDF & unclassified & MF  & MF  & HDF & HDF & unclassified \\ 
 & +$\frac{1}{2}$ & -$\frac{1}{2}$ & +$\frac{1}{2}$ & -$\frac{1}{2}$ & & +$\frac{1}{2}$ & -$\frac{1}{2}$  & +$\frac{1}{2}$ & -$\frac{1}{2}$ &\\
 \midrule
  \#1 (50\% MF) &1 &3 &2 &1 &5 &1 &3 &2 &1 &5  \\ 
\midrule
  \#2 (30\% MF) &2 &1 &3 &3 &2 &3 &1 &4 &1 &2  \\ 
\midrule
  \#3 (70\% MF) &2 &0 &2 &1 &3 &1 &0 &2 &1 &4  \\ 
\midrule
  \#4 (10\% MF) &1 &2 &2 &1 &2 &1 &3 &2 &1 &1  \\ 
\bottomrule
\end{tabular}
\caption{Benchmarking the counting algorithm by comparing count results across four snapshots.}
\label{table:count_comparison}
\end{center}
\end{table}

To automate the process and eliminate counting bias, we implement an algorithm: 
For each defect, we search the area within a radius of 10 pixels (23.4 total pixels) of the defect to compare the number of HDF (magenta) versus MF (green) pixels. If the majority of the area is occupied by green, the defect is classified as ``MF''; if the majority is magenta, it is classified as ``HDF''. If $\Delta n = \dfrac{n_{\mathrm{HDF}} - n_{\mathrm{MF}}}{n_{\mathrm{HDF}} + n_{\mathrm{MF}}}$, which is the normalized difference in number density, if $\Delta n$ falls below a margin threshold $\epsilon$ (the choice of this threshold will be discussed in the next paragraph), then the defect is categorized as ``unclassified''. The results found by the counting algorithm are shown in the second row of Fig. 9.1, and Table \ref{table:count_comparison}, columns 7-11. When choosing $\epsilon$ = {10$^{-4}$}, counting manually and by algorithm agree on 87\% of the counting results for n $\approx$ 50 defects. After verifying the agreement, we apply the algorithm to all frames and all myofibroblast concentrations.

To assess the sensitivity of this classification, we performed a parameter sweep over thresholds and examined how the number of unclassified defects varies (Fig. S9.2). This threshold $\epsilon$ was selected based on the relative intensity between cell-occupied and cell-free regions, with their ratio typically ranging from $10^{-3}$ to $10^{-4}$. While the threshold affects the fraction of unclassified defects, it does not alter the relative counts of each defect type. Therefore, our classification is robust with respect to the choice of threshold. Therefore, we report in the main text the average and standard deviation of the counts for each choice of $\epsilon$.

\begin{figure}[htp!]
    \centering
    \includegraphics[width=0.6\linewidth]{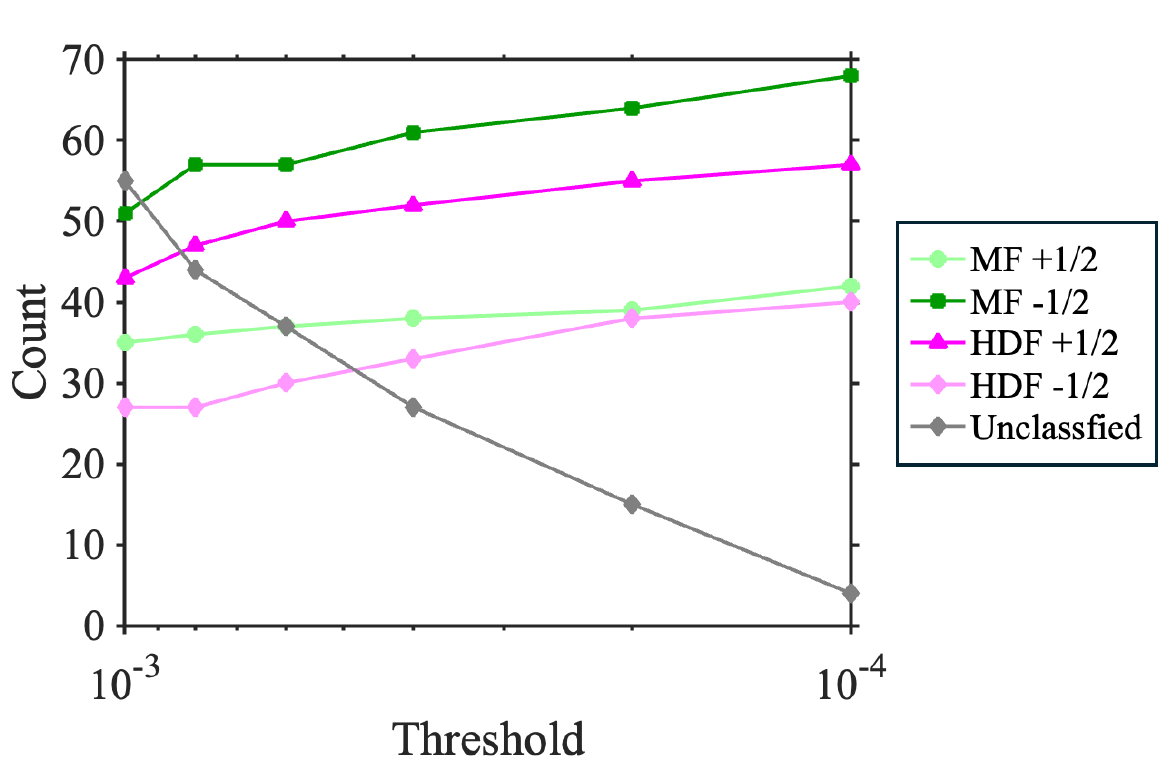}
    \caption*{\textbf{Figure S9.2.} Robustness of the classification by applying different choices of threshold $\epsilon$. }
    \label{fig:threshold}
\end{figure}

\addtocounter{figure}{+1}

\clearpage
\section{Additional results for defect pinning on LCE fiber substrates.}
\begin{figure}[htb!]
    \centering
    \includegraphics[width=0.9\linewidth]{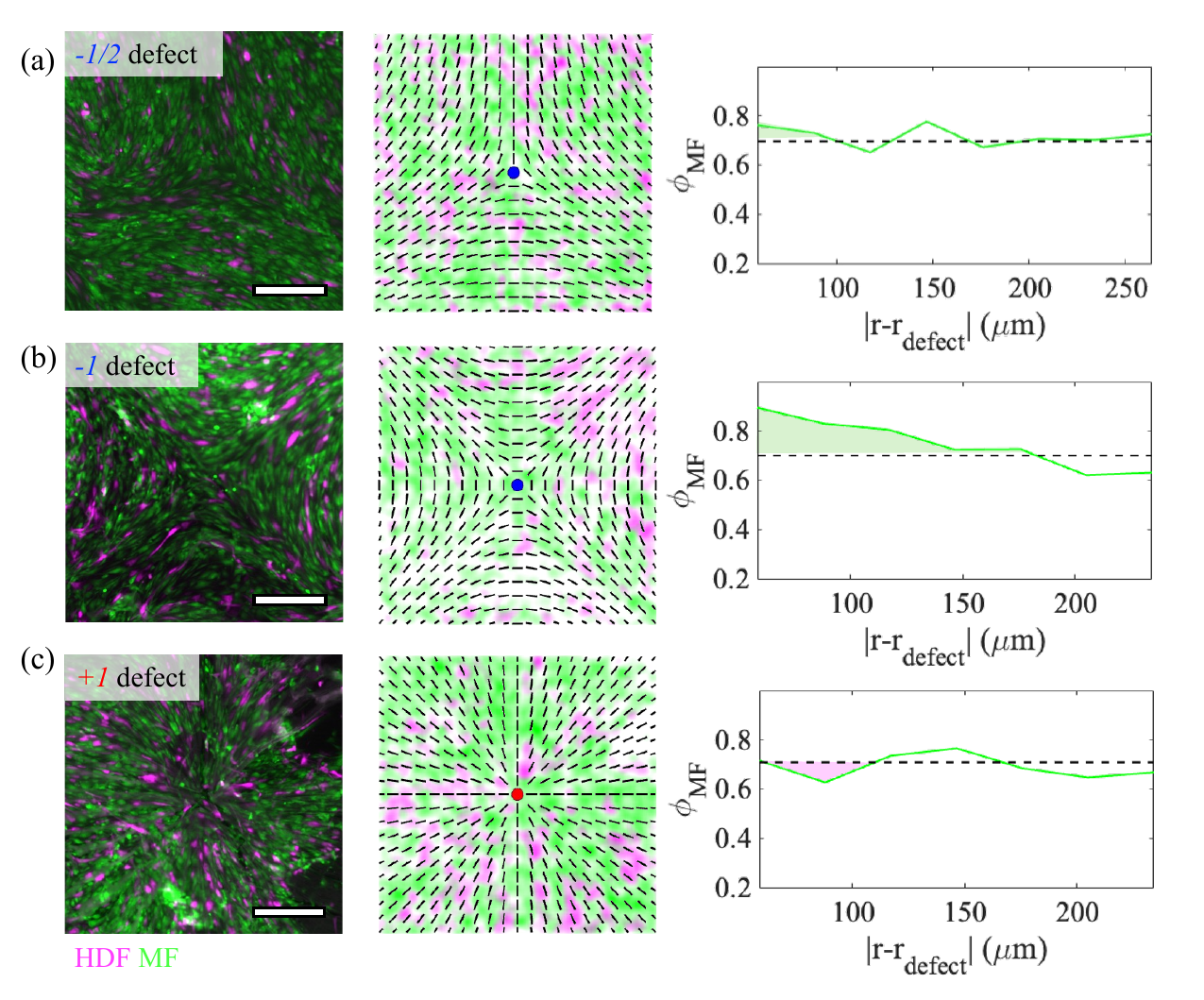}
    \caption{Additional results for defect pinning on LCE fiber substrates. (a) -$\frac{1}{2}$, (b) -1, and (c) +1. Column 1 shows the merged fluorescent micrographs with HDF (magenta) and actin (green). The scale bars are 500 $\mu$m. Column 2 shows the cell density map overlaid with the underlying director field and the defects. Column 3 shows the number density of myofibroblasts within an annulus at a distance $|\mathbf{r}-\mathbf{r}_\mathrm{defect}|$ from the center of the defect. The dashed lines denote the sample average of myofibroblast concentration. The shaded area denotes the range where one phenotype is in excess near the defect core. Note that: cells tend to accumulate at +$\frac{1}{2}$ defects, and because these systems are highly dense, the cell layer is prone to complete delamination from the hydrophobic LCE substrate. }
    \label{fig:more_patterned_defects}
\end{figure}

\clearpage

\section{Yes-associated protein (YAP) immunofluorescence and analysis}
\begin{figure}[htb!]
\begin{center}
    \includegraphics[width=.9\textwidth]{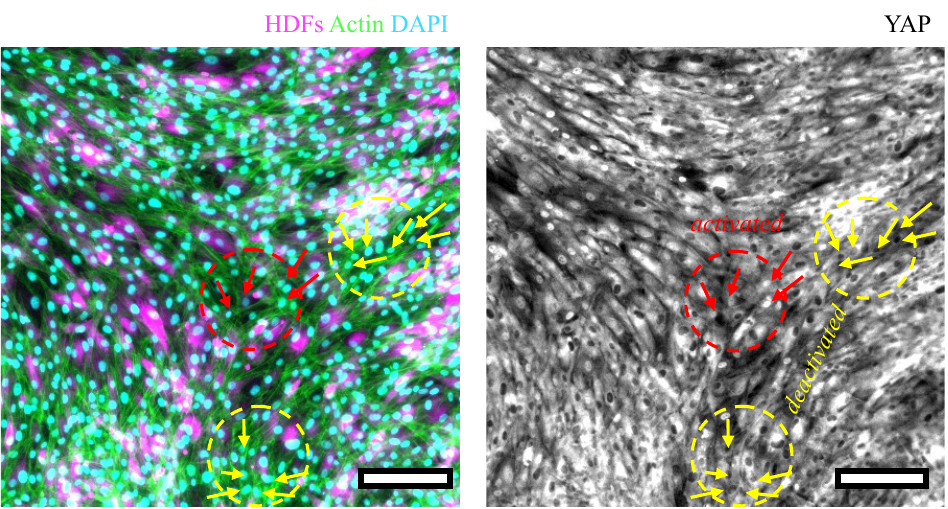}
\end{center}
  \caption{Fluorescence microscopy of a 50:50 fibroblast-myofibroblast mixture seeded on a plain glass substrate, the HDF, actin, and DAPI channels are shown in the image to the left, and YAP staining is shown on the right. The red circle highlights a region at the core of the $-\frac{1}{2}$ defects, and the red arrows indicate myofibroblasts within this region, also visible in the YAP channel. These cells exhibit YAP activation. In contrast, the yellow circles mark two regions away from the defect core where myofibroblasts show YAP deactivation. The scale bars are 200 $\mu$m.}
\label{fig:YAP_minus_half}
\end{figure}

After cell fixation, YAP staining was carried out following procedures in \cite{das2016yap,chen2024disodium}, by first permeating the cells with 0.1\% Triton X-100, and blocking with 1 wt\% BSA and 22.52 mg/mL glycine in PBST (PBS+ 0.1\% Tween 20). Afterwards, the cells were treated with primary YAP antibody at 4$^\circ$C overnight in the dark, followed by secondary IgG. The combined immunofluorescence microscopy is shown below for an isolated $-\frac{1}{2}$ defect in Fig. \ref{fig:YAP_minus_half}. The YAP channel is shown separately to the right. YAP activation is indicated by cells with bright nuclei, whereas YAP deactivation is indicated by cells with bright cytoplasm.

\newpage
\bibliographystyle{unsrt}
\bibliography{refs}

\end{document}